\documentclass{aastex631}

\usepackage{amsmath}
\usepackage{amssymb}
\usepackage{comment}
\usepackage{booktabs}

\newcommand{\m}{M87*}
\newcommand{\s}{Sgr\,A*}

\begin{document}
\defcitealias{Sengupta}{S03}

\title{Effects of Earth's Oblateness on Black Hole Imaging Through Earth-Space and Space-Space VLBI}

\author[0000-0001-8763-4169]{Aditya Tamar}
\affiliation{Department of Physics, National Institute of Technology, Karnataka, Surathkal, Mangalore-575025, India}

\author[0000-0002-3368-1864]{Ben Hudson}
\affiliation{KISPE Space Systems Ltd, Farnborough, United Kingdom}

\author[0000-0002-7179-3816]{Daniel~C.~M.~Palumbo}
\affil{Center for Astrophysics $\vert$ Harvard \& Smithsonian, 60 Garden Street, Cambridge, MA 02138, USA}
\affiliation{Black Hole Initiative at Harvard University, 20 Garden Street, Cambridge, MA 02138, USA}

\begin{abstract}
Earth-based Very Long Baseline Interferometry (VLBI) has made rapid advances in imaging black holes. However, due to the limitations imposed on terrestrial VLBI by the Earth's finite size and turbulent atmosphere, it is imperative to have a space-based component in future VLBI missions. Herein, this paper investigates the effect of the Earth's oblateness, also known as the $J_{2}$ effect, on orbiters in Earth-Space and Space-Space VLBI.
The paper provides an extensive discussion on how the $J_{2}$ effect can directly impact orbit selection for black hole observations and how through informed choices of orbital parameters, the effect can be used to the mission's advantage, a fact that has not been addressed in existing space-VLBI investigations. We provide a  comprehensive study of how the orbital parameters of several current space VLBI proposals will vary specifically due to the $J_{2}$ effect. For black hole accretion flow targets of interest, we have demonstrated how the $J_{2}$ effect leads to modest increase in shorter baseline coverage, filling gaps in the $(u,v)$ plane. Subsequently, we construct a simple analytical formalism that allows isolation of the impact of the $J_{2}$ effect on the $(u,v)$ plane without requiring computationally intensive orbit propagation simulations. By directly constructing $(u,v)$ coverage using the $J_{2}$ affected and invariant equations of motion, we obtain distinct coverage patterns for \m{} and \s{}  that show extremely dense coverage on short baselines as well as long term orbital  stability on longer baselines. \\
\end{abstract}


\section{Introduction} \label{sec:intro}
The field of radio astronomy, in particular very long baseline interferometry (VLBI), has made rapid advances in the field of black hole imaging, stimulated primarily by the observations of the Event Horizon Telescope (EHT) collaboration. The images of \m{} \citep{EHT:2019:1} and \s{} \citep{collaboration_first_2022} provide a novel means to access the strong-field regime of black holes and enable direct probes of the magnetic field structures in the inner accretion flow through the polarization of synchrotron radiation \citep{2021ApJ...910L..12E, EHT:2019:8}. Observations with lower frequency arrays such as the GMVA continue to improve our understanding of black hole accretion and jet launching \citep{Lu:2023} as well. \\
However, it is important to note that the EHT, and its immediate successor, the next generation EHT (ngEHT) \citep{Johnson:2023,Doeleman:2023,Ayzenberg:2023} are Earth-based networks of observing stations. This has direct physical implications on the scientific value of the observed data. For example, the maximum length of a baseline that can be obtained is constrained by the diameter of the Earth.\citep{collaboration_first_2019} and the observing frequency has to be chosen so as to ameliorate the effects of the Earth's atmosphere \citep{EHT:2019:3,Raymond:2021}. \\
It is therefore apparent that in order to continue improving the resolution and the Fourier plane (hereafter $(u,v)$) coverage, future VLBI missions targeting black hole astrophysics must have a space-based component. Indeed, there have been several recent mission concept studies in this domain \citep{Andrianov:2021,Johnson:2021,Likha2022,Rudnit:2023,Kudri2021,roelofs_ngeht_2023,Kudri2021,Peter:2022,Trippe2023,Hudson_2023}. The space missions can be a single orbiter working in tandem with a ground based station to produce very long baselines, in turn leading to very high angular resolution. However, in this case one would still have to deal with effects of Earth's atmosphere. The other alternative is of space-only VLBI that contains several orbiters forming baselines with one another, wherein deciding an appropriate formation geometry is a key task, encompassing both astrodynamics and VLBI site optimisation. Herein, one would eliminate any corrupting effects due to the Earth's atmosphere. \\
In this paper, we provide a detailed investigation of how one realistic astrodynamical effect, namely the effect of Earth's oblateness on an orbiter's motion (the so-called $J_{2}$ effect), impacts the $(u,v)$ coverage for observing \m{} and \s{} using both Earth-Space and Space-Space VLBI. The choice of these sources is to develop a linear trajectory of improvement over the existing observations of the EHT. Now, it is a reasonable question to ask why specifically the $J_{2}$ effect is being studied when sophisticated orbit propagation models are already being used for space VLBI mission studies. These models include not just the $J_{2}$ effect, but also effects of air drag, dynamics of the upper atmosphere and so on \citep{Vallado:2006}. The reason for specifically investigating the $J_{2}$ effect is due to its peculiar nature wherein by an informed choice of related orbital parameters, the effect can be used to one's \textit{advantage}, as compared to being a ``corrupting" influence on our observations. In particular, choices of orbital parameters informed by analysing the $J_{2}$ effect will lead to a significant impact for calculations related to the fuel budget of the mission, a non-trivial factor in mission design. \\
Keeping all of these factors in mind, the paper studies the $J_{2}$ effect on the $(u,v)$ coverage using analytical tools developed in astrodynamical literature that model this effect for the orbiter's motion. Since we are directly using the analytic expressions, the framework is much less computationally expensive than sophisticated orbit propagation models, while also providing insights into how such physical effects can impact black hole observations through a space-VLBI component. 
The layout of the paper is as follows: In Section \ref{sec2} we provide an accessible introduction to the $J_{2}$ effect, its effect on orbital parameters and how the effect can be used to one's advantage when selecting orbits to maximise scientific output from the coverage of the $(u,v)$ in observing \m{} and \s{}. In Section \ref{sec3} we provide a detailed discussion of how the orbital parameters of existing VLBI missions with a space-based components are affected by the $J_{2}$ effect. In addition, a brief discussion is provided on the effect's impact on the visibility domain signature of a photon ring modelled as an infinitesimally thin ring of unit flux. In Section \ref{sec4} a novel analytic formalism is developed that describes the $(u,v)$ coverage in terms of equations of motion of two space-based orbiters in relative motion of one another, namely the chief and the deputy. In Section \ref{sec5} the relative motion is studied without incorporating the $J_{2}$ effect and the corresponding features of the baseline coverage are studied for both \m{} and \s{}. In Section \ref{sec6}, the $J_{2}$ effect is incorporated through a linearisation scheme in the equations of motion and the corresponding $(u,v)$ coverage for the same two black holes is studied. In Section \ref{sec7}, the equations of motion that produce orbits which are invariant under the $J_{2}$ effect are used and the bounded relative motion is observed in the $(u,v)$ plane as well.  Finally, Section \ref{sec8} provides the conclusion, as well as some potential avenues for future work. For the benefit of the reader, a tabulation of some of the terminology used in the paper has been provided in Table \ref{tab:terminology} in the Appendix.

\section[Incorporating Astrodynamics Considerations into VLBI: Effect of Earth's Gravitational Field via the J2 Perturbation]{Incorporating Astrodynamics Considerations into VLBI: Effect of Earth's Gravitational Field via the $J_{2}$ Perturbation}\label{sec2}
The Earth is essentially an oblate spheroid, and is flatter at the poles and bulging out at the equator \citep{Schaub,Handbook}. As a consequence of this, the gravitational attraction towards a body orbiting the Earth is not directed towards the Earth's centre of mass, as would be expected from classical Newtonian theory of gravitation. A common approach to model this effect on an orbiter is by expressing the gravitational potential in terms of spherical harmonics. If $r,\phi,\lambda$ are the satellite's orbital radius, latitude, and longitude respectively, then the potential can be written as: 
\begin{gather}
    V(r)=\frac{\mu}{r}\sum_{l=0}^{\infty}\Bigg(\frac{a_{e}}{r}\Bigg)^{l}\sum_{m=0}^{l}P_{lm}(\sin\phi)(C_{lm}\cos m\lambda+S_{lm}\sin m\lambda),
\end{gather}
wherein $\mu$ is the product of Newton's gravitational constant and mass of Earth and $a_{e}$ is the semi-major axis of Earth's ellipsoid shape. The functions $P_{lm}(..)$  are the associated Legendre functions of degree $l$ and order $m$ and red$C_{lm}$, $S_{lm}$ are spherical harmonic coefficients. Extremely accurate satellite measurements of these coefficients have been incorporated into the Joint Gravity Model 3 (JGM-3) developed by NASA \citep{Tapley:1996}.\\
To understand the separate effects of the harmonics, one often splits this equation into three parts:
\begin{gather}
    V=V_{0}+V_{1}+V_{2},
\end{gather}
where
\begin{gather}
    V_{0}=\frac{\mu}{r}, \\
    V_{1}=\frac{\mu}{r}\sum_{l=1}^{\infty}\Bigg(\frac{a_{e}}{r}\Bigg)^{l}P_{l0}(\sin\phi)C_{l0},\\
    V_{2}=\frac{\mu}{r}\sum_{l=1}^{\infty}\Bigg(\frac{a_{e}}{r}\Bigg)^{l}\sum_{m=1}^{l}P_{lm}(\sin\phi)(C_{lm}\cos m\lambda+S_{lm}\sin m\lambda).
\end{gather}
Here, $V_{0}$ is the classical potential when Earth is treated as a point mass. The term $V_{1}$ corresponds to the part of the potential which does not have longitudinal dependence and hence has $m=0$. By defining
\begin{gather}
    J_{l}\equiv -C_{l0},
\end{gather}
the ``zonal" part of the potential is now written as:
\begin{gather}
    V_{1}=-\frac{\mu}{r}\sum_{l=1}^{\infty}\Bigg(\frac{a_{e}}{r}\Bigg)^{l}P_{l0}(\sin\phi)J_{l}. \label{eq:zonhar}
\end{gather}
The remaining part $V_{2}$ is dependent on the longitude.\\
For our purposes, we focus on Equation \ref{eq:zonhar}. It is the degree 2 zonal term, defined as $J_{2}$ in the equation, that models the contribution due to Earth's oblateness and hence the change in dynamics arising due to it are sometimes known as the $J_{2}$ perturbation effect \citep{KH:1958}. For this paper, we shall use the standard value of the $J_{2}$ coefficient \citep[][hereafter \citetalias{Sengupta}]{Sengupta}
\begin{gather}
    J_{2}=1.082629\times10^{-3}.
\end{gather}
 As far as the effects of the $J_{2}$ perturbation are concerned, it is the dominant perturbing influence for orbiters in Low Earth Orbits (LEOs). Indeed, the Orbiter-Earth system can be modelled as a Keplerian 2-body problem and the dominant contribution of the $J_{2}$ terms can be investigated using Earth models such as the JGM-3. For LEO deployment, incorporation of $J_{2}$ effects via the JGM-3 model has a relatively better performance than the more widely used Simplified General Perturbation-4 (SGP4) model \citep{Morales:2019}. \\
In the context of Earth-space VLBI, the motivation for studying its impact arises from the fact that several recent discussions of future missions have discussed placing stations in LEOs \citep{Palumbo2019}. At altitudes of around 800 km (which is in the LEO range), the $J_{2}$ perturbation is the dominant effect when compared to other realistic considerations such as atmospheric drag, solar radiation pressure \citep{Alessi:2018} and other electromagnetic effects \citep{Marsden2001}. For some related work on the non-trivial importance of $J_{2}$ on relative orbit motion, see \citet{Sam2002} and for efforts on modelling orbits invariant to $J_{2}$ effects, see \citet{Schaub:2001,Schaub,Lee:2022}.\\

\subsection[Effect of J2 perturbation on orbital parameters]{Effect of $J_{2}$ perturbation on orbital parameters}
For a given orbiter in space, there are 6 Keplerian orbital elements that can describe its behaviour: semimajor axis $a$, eccentricity $e$, inclination $i$, right ascension $\Omega$, argument of perigee $\omega$ and mean anomaly $M$. Due to the $J_{2}$ perturbation, over a large number of orbits, the time evolution of these elements is affected as follows:
\begin{gather}
\dot{a}=0,\quad \dot{e}=0,\quad \dot{i}=0, \label{eq:aeij2}\\
\dot{\Omega}=-\frac{3}{2}nJ_{2}\Bigg(\frac{R_{e}}{p}\Bigg)^{2}\cos i, \label{eq:Omj2}\\
\dot{\omega}=\frac{3}{4}nJ_{2}\Bigg(\frac{R_{e}}{p}\Bigg)^{2}(5\cos^{2}i-1) \label{eq:omj2},\\
\dot{M}=n[1+\frac{3}{4}\sqrt{1-e^{2}}J_{2}\Bigg(\frac{R_{e}}{p}\Bigg)^{2}(3\cos^{2}i-1)], \label{eq:Mj2}
\end{gather} 
where
\begin{equation}
    n=\sqrt{\frac{\mu}{a^{3}}},\quad p=a\sqrt{1-e^{2}} \label{eq:meanmotion}.
\end{equation}
 and the ``overdot" represents the time derivative of the corresponding parameters. Now the longitude $\Omega$, inclination $i$ and latitude $\theta$ form a $3-1-3$ Euler angle system \citepalias{Sengupta}. Moreover, considering long-term variation in orbital parameters over a large number of orbits, only effects of $\dot{\omega}$ are felt so we approximate \citepalias{Sengupta}
\begin{equation}
    \dot{\theta}\approx\dot{\omega}, \label{eq:vartheta}
\end{equation}
thereby ignoring the effects of the harmonics of the true anomaly $f$. This approximation is valid for mission cycles of space VLBI missions such as Millimetron, which has an expected 10 year cycle of operation \citep{Lazio2020}.\\
To get a better understanding on how the orbital elements evolve over time, we can integrate the Equations \ref{eq:Omj2}, \ref{eq:omj2} and \ref{eq:Mj2} assuming initial values $\Omega_{0}$, $\omega_{0}$ and $M_{0}$ respectively. Note that due to Equation \ref{eq:aeij2}, the inclination $i$ in this framework does not have time dependence. One obtains the following expressions:
\begin{gather}
    \Omega(t)=\Bigg(-\frac{3}{2}nJ_{2}\Bigg(\frac{R_{e}}{p}\Bigg)^{2}\cos i\Bigg)t+\Omega_{0},\\
    \omega(t)=\Bigg(\frac{3}{4}nJ_{2}\Bigg(\frac{R_{e}}{p}\Bigg)^{2}(5\cos^{2}i-1)\Bigg)t+\omega_{0},\\
    M(t)=\Bigg(n\Bigg[1+\frac{3}{4}\sqrt{1-e^{2}}J_{2}\Bigg(\frac{R_{e}}{p}\Bigg)^{2}(3\cos^{2}i-1)\Bigg]\Bigg)t+M_{0}. \label{eq:varorb}
\end{gather}
For future reference, we also note the approximate relation between true anomaly $\nu$ and mean anomaly:
\begin{gather}
    \nu=M+(2e-\frac{1}{4}e^{2})\sin M +\frac{5}{4}e^{2}\sin 2M +\frac{13}{12}e^{3}\sin 3M +\mathcal{O}(e^{4}).
\end{gather}
Upon differentiating this relation, and assuming $\dot{e}=0$ as in the case for $J_{2}$'s effect, we get:
\begin{gather}
    \dot{\nu}_{J2}=\dot{M}\Bigg[1+(2e-\frac{1}{4}e^{2})\cos M+\frac{5}{2}e^{2}\cos 2M+\frac{13}{4}e^{3}\cos 3M\Bigg].
\end{gather}
For the choice of orbits given in Table \ref{tab:orbits} (justification for which is provided in the following section), the variation of $\Omega, \omega$ for a time period of 1 day and 1 month are tabulated in Table \ref{tab:orbj2}. The $J_{2}$ effect is apparent here. For example, the choice for $\omega$ for the Equatorial HEO is chosen to maximise the coverage of \m{} but the parameter has variability even over a single day, and it only increases with time. This causes variation in the coverage of \m{} that can be achieved, as will be demonstrated in subsequent sections.
As another example, the effect of the so-called ``critical/magic inclination" of $i=63.4^{\circ}$ for the Molniya orbit is apparent since there is no time variation in $\omega$. Such an inclination has been known in astrodynamics literature (for example, see \citet{Sabatini:2008}) for some time, with Molniya orbits being used for satellite orbits since the 1960's \cite{Allan:1971}. The numerical value of this inclination is obtained as follows:
\begin{gather}
    \dot{\omega}=0\Rightarrow(5\cos^{2}-1)=0\Rightarrow i=\cos^{-1}(1/\sqrt{5})=63.4^{\circ}. \label{eq:molniyamimj2}
\end{gather}
Therefore, suitable modifications/extensions of the Molniya orbit for maximising the coverage of a particular source (just like we did for Equatorial HEO) can be used to counteract the variation in source coverage that might arise from the $J_{2}$ effect on orbital parameter $\omega$.

\subsection{Orbit Selection}
As with many aspects of space system design, the orbit is often a trade-off between what is most desirable to meet the scientific objectives and the feasibility from a mission analysis perspective. Orbit selection will affect many aspects of the mission including power generation, thermal environment, communications and radiation exposure, to name a few. The effects on these factors must be carefully balanced through orbit selection to ensure the system can successfully meet the primary objectives and operate safely in the space environment.

For VLBI, orbit selection is primarily a geometrical consideration to achieve sufficient $(u,v)$ coverage and resolution when observing the target astronomical sources. Considering a single space telescope observing in collaboration with ground-based stations, angular resolution
will be driven by the maximum altitude of the orbit as increasing the baseline length is required to improve resolution. For VLBI imaging dense $(u,v)$ coverage is required to increase the fidelity of the generated images. This is achieved by varying the baseline length throughout observations, therefore operating at a range of altitudes for a space-based system. Rapid $(u,v)$ coverage is also desirable as the radio emitting environment of black holes is highly variable over short timescales. To capture the dynamic behaviour of supermassive black holes, coverage should be achieved in the shortest observation time possible (see analysis in \citet{roelofs_ngeht_2023}). For an orbiting antenna, this encourages the selection of an orbit with a short time period (i.e. high orbital velocity) to sample a range of $(u,v)$ points in as short a time as possible.

\begin{figure}
    \centering
    \includegraphics[width=0.4\columnwidth]{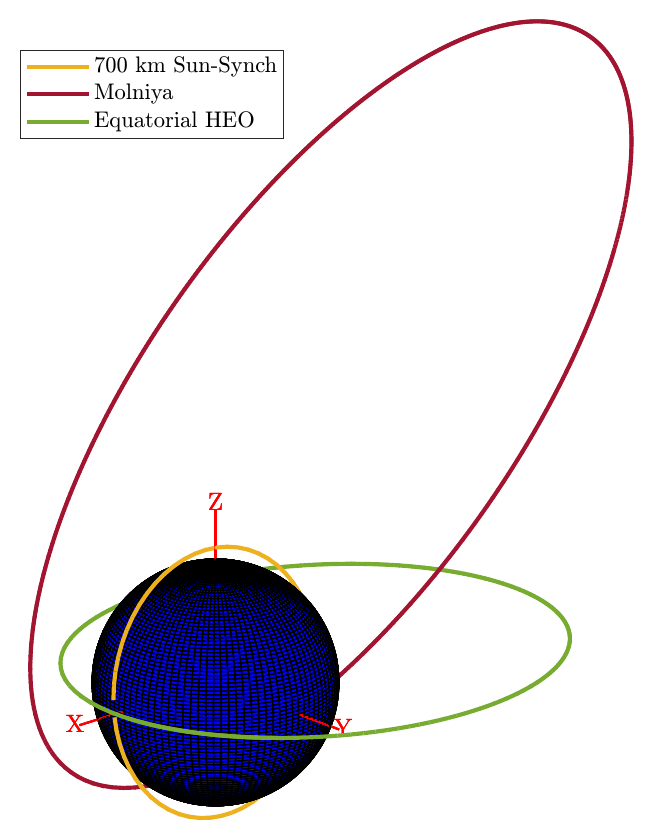}
    \caption{Analysed orbit configurations plotted in the ECI reference frame.}
        \label{f:orbitPlot}
\end{figure}

For this study a number of commonly used Earth orbits are evaluated in terms of their benefits from a VLBI and mission analysis perspective. The subset of orbits selected for analysis is by no means comprehensive but provide a good example of the trade-offs which must be resolved and the effect of the $J_{2}$ perturbation on the mission and subsequently, the VLBI observations. Table \ref{tab:orbits} contains the three analysed orbits with their respective Keplerian elements. Fig. \ref{f:orbitPlot} depicts the configurations in the Earth Centered Inertial (ECI) frame. The ECI frame $(\hat{X},\hat{Y},\hat{Z})$ is such that $(\hat{X},\hat{Y}$) span the equatorial plane of the Earth, $\hat{Z}$ is along the North pole and $\hat{X}$ is along the vernal equinox . The origin of the system is at the Earth's centre. This is analogous to the right-handed Cartesian co-ordinate system used in VLBI for specifying positions of antennas in an array \citep{Thom}.\\
\begin{table}
\begin{center}
$\begin{array}{lccc}
    \begin{tabular}{m{5cm} m{3cm} m{3cm} m{3cm}}
        \toprule
        Orbit Elements & $700$ km Sun-Sync & Equatorial HEO &  Molniya Orbit \\
        \midrule
        Semi-major axis (\(a\)) [km]	& 7078  &   14000  &   26600 \\
        Eccentricity (\(e\))	& 0  &   0.50   &   0.74 \\
        Inclination (\(i\)) [\(^\circ\)]	& 97.4  &   0.0   &   63.4 \\
        Right Ascension (\(\Omega\)) [\(^\circ\)]	& 0*  &   0   &   0 \\
        Argument of Perigee (\(\omega\)) [\(^\circ\)] & 0 &   187   &   270 \\
        Period [hours] & 1.7 & 4.6 & 12.0 \\
        \bottomrule
    \end{tabular}
    
\end{array}$
\end{center}
\caption{Keplerian elements of three prospective Earth orbits for a space-based interferometer. *Right ascension of a Sun-synchronous orbit must be selected based on launch date and relative Sun position.}
\label{tab:orbits}
\end{table}
Now, depending on the specific objectives of the mission, the $J_{2}$ perturbation on right ascension and argument of perigee can either by beneficial or detrimental. For VLBI, careful selection of orbital elements could result in the nodal and apsidal precession being advantageous for the observation of supermassive black hole targets.

\begin{table}
\begin{center}
$\begin{array}{ccccc}
    \begin{tabular}{m{4cm} m{3cm} m{3cm} m{3cm} m{3cm}}
        \toprule
        Orbit & Orbital Parameter & Initial Value &  Value after 1 day & Value after 1 month \\
        \midrule
        $700$ km Sun-Sync & $\Omega$  &   $0^{\circ}$  &   $0.32043^{\circ}$ & $9.7464^{\circ}$ \\
        	$700$ km Sun-Sync & $\omega$  &   $0^{\circ}$   &   $-1.14077^{\circ}$ & $-34.6985^{\circ}$ \\
        Equatorial HEO	& $\Omega$  &   $0^{\circ}$   &   $-0.84788^{\circ}$ & $-25.7898^{\circ}$ \\
        Equatorial HEO	& $\omega$  &   $187^{\circ}$   &   $188.696^{\circ}$ & $238.58^{\circ}$ \\
        Molniya & $\Omega$ &   $0^{\circ}$   &   $-0.17416^{\circ}$ & $-5.29747^{\circ}$ \\
        Molniya & $\omega$ &   $270^{\circ}$   &   $270^{\circ}$ & $270.002^{\circ}$
    \end{tabular}
    
    \end{array}$
\end{center}
\caption{Variation of orbital elements $\Omega$ and $\omega$ due to the $J_{2}$ effect after 1 day and 1 month}
\label{tab:orbj2}
\end{table}

A Sun-synchronous orbit is a good example of a LEO that makes use of the $J_{2}$ perturbation on the right ascension of the ascending node. Many Earth-observing satellites observing at optical wavelengths require consistent lighting of specific targets. Sun-synchronous orbits make use of the nodal precession due to the $J_{2}$ perturbation to rotate the right ascension of the orbit so that the position of the Sun with respect to the orbital plane is constant throughout the year. The inclination and semi-major axis of a Sun-synchronous orbit are selected such that the nodal precession is equal to the rate of the Earth's rotation about the Sun. For VLBI observations this is highly beneficial. Due to the stringent sensitivity requirements on VLBI interferometers, it is highly likely that space-based VLBI systems will require cooling of the receiver electronics to very low temperatures, depending on the frequency selected, as discussed in \citet{gurvits_science_2022}. Like many space telescopes (e.g. James Webb), this may require all observations to be conducted away from the solar direction to simplify thermal control onboard. Maintaining the Sun in a constant position with respect to the orbital plane would simplify operations of the spacecraft drastically. A 700~km altitude has been selected for this orbit as this keeps the spacecraft outside of the Inner Van Allen Belt which begins to affect spacecraft electronics at higher altitudes. One notes that the use of a Sun-synchronous orbit for VLBI observations is one example of how careful orbit selection can make use of the orbit precession due to the Earth's oblateness in a beneficial way. 

An equatorial, Highly Elliptic Orbit (HEO) has been selected to demonstrate the apsidal precession effects of the $J_{2}$ perturbation. Having an inclination of 0\(^\circ\) maximises the rate of change of the argument of perigee. The semi-major axis and eccentricity have been derived by selecting a perigee altitude of 700~km (the same as the Sun-synch) and an apogee of 21000~km (half of the Molniya) to provide a variation in $(u,v)$ coverage from the other two orbits. The argument of perigee has been calculated to maximise coverage of \m{}. As with the Molniya orbit, the increased apogee altitude compared to the Sun-sync provides a longer baseline to a ground-based station and thus, finer angular resolution. Unlike the Molniya, the period of this orbit is only \(\sim\)4.5 hours providing more rapid $(u,v)$ coverage as the spacecraft completes a full revolution of the Earth in almost a third of the time.

Molniya orbits were first designed to provide long periods of coverage over high latitudes, as opposed to the coverage typically offered by traditional communications satellites in Geostationary Earth Orbit (GEO). In order to achieve this, the orbit has a high eccentricity to increase the period of time spent over the target region. As has been discussed, apsidal precession due to the $J_{2}$ perturbation results in rotation of the argument of perigee over time. To combat this, the Molniya orbit makes use of a 63.4\(^\circ\) inclination which results in zero apsidal precession. This is generally referred to as a frozen orbit as some perturbation effects have been cancelled out through the orbital element selection. This feature, along with the orbit's large altitude variation (hence baseline variation), makes it of particular interest for this study.

\begin{figure}
     \centering
     \includegraphics[width=0.4\textwidth]{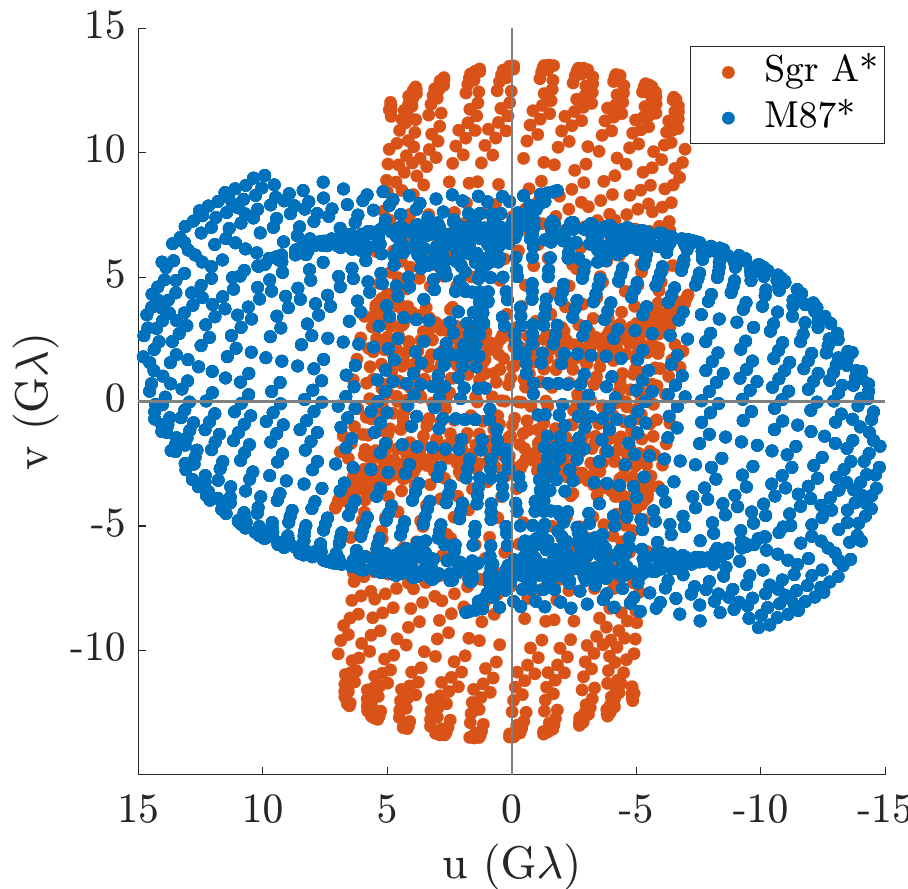}
     \caption{$(u,v)$ coverage achieved by Sun-synchronous orbit configuration with the Large Millimeter Telescope (LMT), observation at 345~GHz.}
     \label{f:SunSynchUV}
\end{figure}

\begin{figure}
     \centering
     \includegraphics[width=0.4\textwidth]{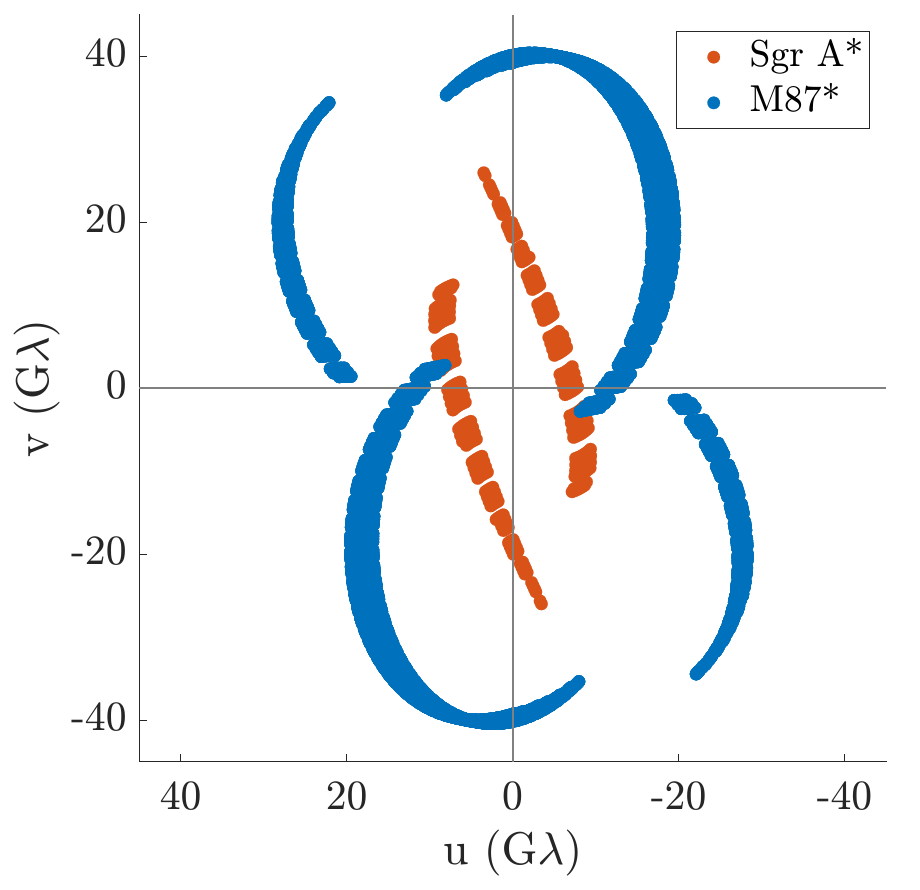}
     \caption{$(u,v)$ coverage achieved by Molniya orbit configuration with the Large Millimeter Telescope (LMT), observation at 345~GHz.}
     \label{f:MolniyaUV}
\end{figure}

\begin{figure}
     \centering
     \includegraphics[width=0.8\textwidth]{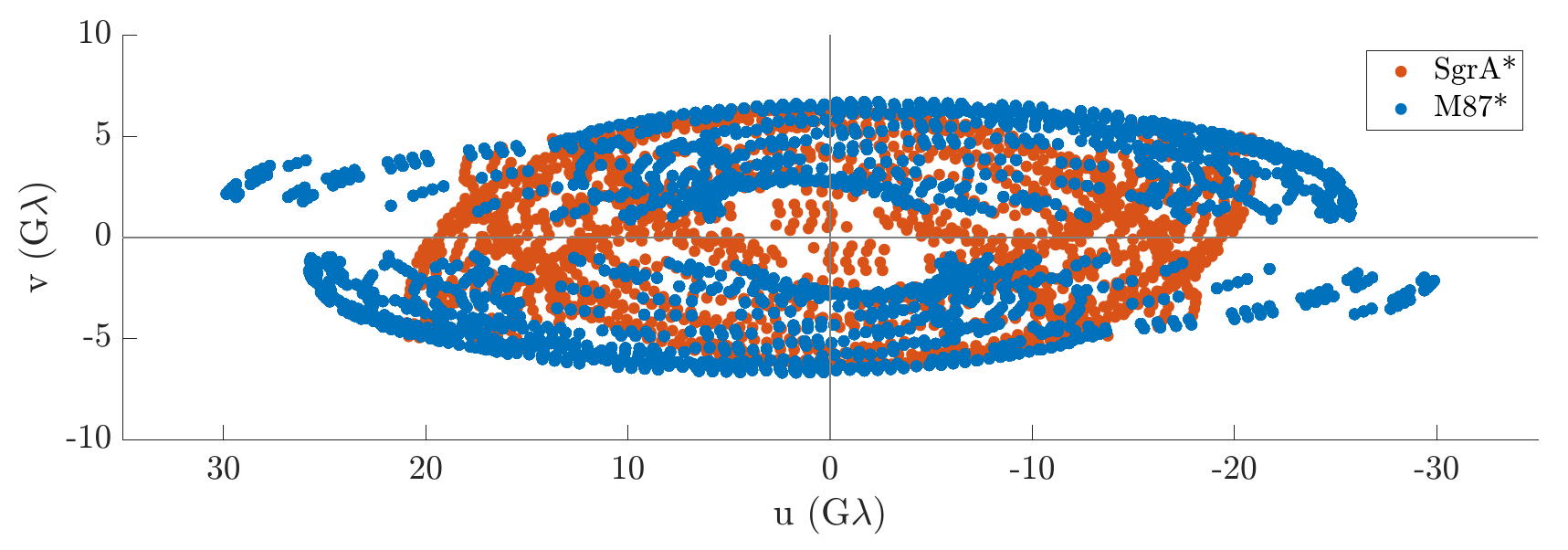}
     \caption{$(u,v)$ coverage achieved by equatorial HEO configuration with the Large Millimeter Telescope (LMT), observation at 345~GHz.}
     \label{f:EquatorialUV}
\end{figure}

Fig. \ref{f:SunSynchUV}, Fig. \ref{f:MolniyaUV} and \ref{f:EquatorialUV} depicts the $(u,v)$ coverage achieved by each configuration for 7 day observations of \m{} and \s{}. Observations are conducted at 345~GHz, with a single ground station, in this case the Large Millimeter Telescope (LMT) in Mexico. Observation could however be conducted with any ground-based antenna and still produce the same high-level $(u,v)$ coverage features which are described below. 
As stated in \citet{roelofs_ngeht_2023}, this is the target frequency of the next generation EHT (ngEHT). For these simulations, each scan is conducted for 5 minutes, the average length of measurements conducted by the EHT in the observations of \m{} (see \citet{collaboration_first_2019}). The instrument duty cycle is 10 minutes, providing 5 minutes between the end of one scan and the start of the next. The exact duty cycle of a space-based instrument will depend on many factors such as power requirements, data processing and storage, thermal conditions and observation of calibration sources. However, the general pattern of the $(u,v)$ coverage remains consistent regardless of the duty cycle and it is simply the density of the coverage that will vary.

The variations in the orbits selected can be clearly seen in their respective $(u,v)$ coverages of \m{} and \s{}, shown in Fig. \ref{f:SunSynchUV}, Fig. \ref{f:MolniyaUV} and \ref{f:EquatorialUV}. As expected, the finest resolution of 5.1~\(\mu\)as is achieved by the Molniya orbit configuration as it has the largest apogee of the three designs. However, due to the long period of this orbit (12 hours), the $(u,v)$ coverage is very sparse compared to the Sun-synchronous and equatorial alternatives which complete 7.5 and 2.5 more revolutions about the Earth, respectively, in the same time. As \m{} is at a fairly low declination, the $(u,v)$ coverage of the equatorial HEO is less varied in the $v$-plane than the  alternative, inclined orbits.

All three configurations clearly demonstrate the benefits of space-based VLBI stations. The improvement in angular resolution is considerable compared to the 25~\(\mu\)as achieved by the EHT in its \m{} and \s{} observations (see \citet{collaboration_first_2019}, \citet{collaboration_first_2022}). The Sun-synchronous orbiter produces a far denser $(u,v)$ coverage and a resolution of 20.8~\(\mu\)as for M87*. The increased altitude of the equatorial HEO exhibits a resolution of 6.9~\(\mu\)as, an improvement on the EHT by almost a factor of 5. These results were achieved by a simple interferometer of only two elements, the orbiter and the ground-based LMT. Denser coverage would be possible for observations with many more ground stations, even for the sparsely populated Molniya orbiter $(u,v)$ plot. The long baselines achieved by the equatorial HEO and the Molniya orbit configurations also introduce the possibility of probing the first order photon ring of \m{}. \citet{johnson_universal_2020} showed that that the interferometric signature of the black hole is dominated by the photon ring contribution beyond \(\sim\)20~G\(\lambda\) and that the sub-mm interferometric signatures of \s{} and \m{} should be dominated by the photon ring contribution. Detection of the signature on very long baselines can allow one to estimate the angular diameter of the photon rings, from which uniquely robust and accurate tests of strong gravity and general relativity can be performed. 

As is demonstrated in Fig. \ref{f:SunSynchUV}, Fig. \ref{f:MolniyaUV} and \ref{f:EquatorialUV}, the $(u,v)$ coverage of a radio source is dependent on the geometry of the orbit design. It is therefore possible to optimise the orbit configuration to maximise visibility and coverage of a certain source. The investment required to fund a future space-based VLBI mission will be significant and therefore the system will need to be highly versatile to justify the cost. Optimisation for observation of a single source will not be desirable. Changing the shape and orientation of an orbit can be an expensive orbit transfer. Conducting plane change maneuvers for varying inclination and right ascension are particularly demanding and often completely impractical in terms of the required propellant. The relative right ascension of the source with respect to the orbital plane will vary throughout the year due to the nodal and apsidal precession caused by J2. This natural variation can be used to change the visibility that the space-based element of the interferometer has of different sources.

 \begin{figure}[t]
    \centering
    \includegraphics[width=0.7\columnwidth]{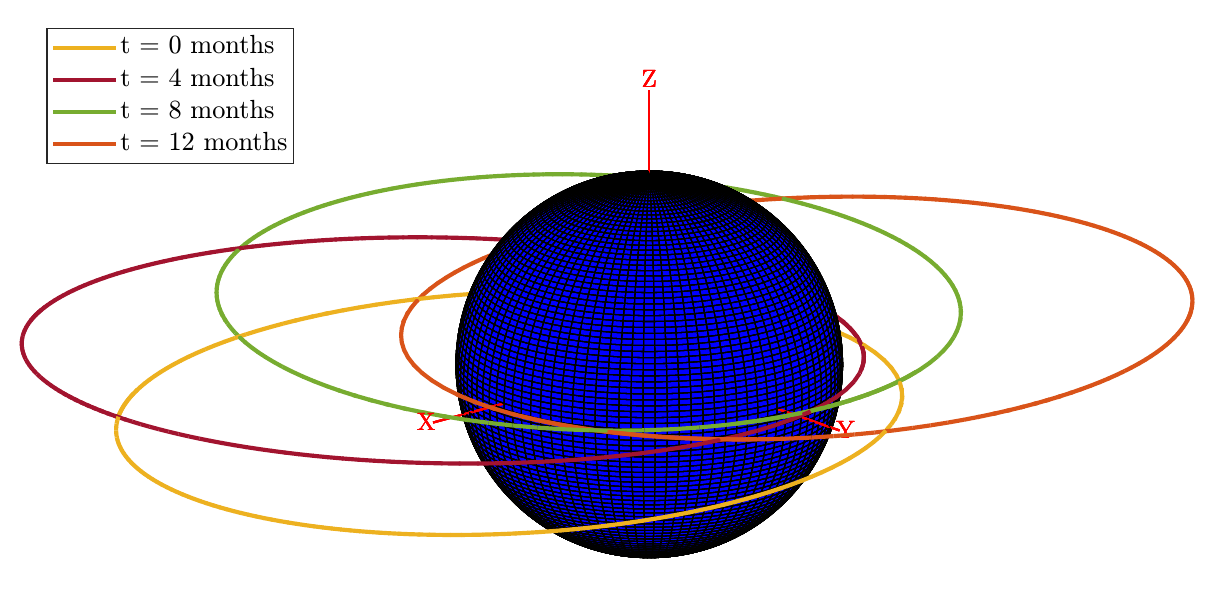}
    \caption{Equatorial HEO precession due to the $J_{2}$ perturbation across a 12 month period.}
    \label{f:equatorialPrec}
\end{figure}

Therefore, precession due to $J_{2}$ should be considered when designing orbit configurations to optimise coverage of certain sources. The equatorial HEO was selected to maximise the rate of apsidal precession to illustrate this approach. Fig. \ref{f:equatorialPrec} depicts the precession of this orbit across a 12~month period. The argument of perigee changes by 183~\(^\circ\) due to the $J_{2}$ perturbation in a single year. The initial argument of perigee was selected to maximise the baseline length achieved for observations of \m{}. The resolution of \s{} that would be achieved is \(\sim\)70\% of that for \m{}. In this configuration, 8~months after the initial conditions the effect of the $J_{2}$ perturbation results in the resolution achieved during observations of \s{} improving from 9.8~\(\mu\)as to 8.0~\(\mu\)as.

This highlights one of the disadvantages of the Molniya configuration and other frozen orbit types. By cancelling out the $J_{2}$ perturbation on right ascension and argument of perigee, the natural precession effects cannot be used to vary coverage of different sources. However, as a consequence, sometimes months would pass between the optimal observation times for different sources. It is however likely that this would be acceptable for a mission that would have a lifetime of at least 5~years. Scheduling with ground based arrays would need to be carefully conducted such that observations could take place at the optimal time. The orbit configuration of a space-based station can therefore be designed such that the natural perturbations due to $J_{2}$ are advantageous for the observation of radio sources and the operation of the spacecraft platform itself. However, ignoring the apsidal and nodal precession effects would likely result in an unfavourable geometry for certain sources that could have otherwise been avoided.

\section[Investigating J2 effects in Existing Earth-Space and Space-VLBI proposals]{Investigating $J_{2}$ effects in Existing Earth-Space and Space-VLBI proposals}\label{sec3}
Recently, there have been several mission concept studies in literature that have investigated the scientific merits of having a space-based VLBI station \citep{Fromm2021,roelofs_ngeht_2023,Rudnit2022,Likha2022,Trippe2023,Peter:2022}. These merits are intimately tied to the ``cartography" of the $(u,v)$ plots, with larger baselines implying higher angular resolution and shorter baselines improving the density of the $(u,v)$ plane. Now, existing literature has explored in detail the scientific payoff of the increased coverage of the $(u,v)$ plane with space VLBI. However, an extended discussion of how realistic astrodynamical effects acting on the orbiter can affect the $(u,v)$ coverage has been lacking in these studies. Therefore, we now investigate how the $J_{2}$ perturbation specifically affects the orbital elements for the said mission concept studies.  The focus is on those studies in which the relevant orbital parameters affected by $J_{2}$ have been explicitly provided.\\
Firstly, the parameters in \citet{Fromm2021} are investigated, wherein the authors devised an optimisation algorithm to observe \s{} over a 12h time period. The proposal involved a single space-based orbiter in a Medium Earth Orbit (MEO). The orbital parameters of the proposal, along with the change in values due to $J_{2}$ are given in Table \ref{tab:orbj2fromm}.
\begin{table}
\begin{center}
$\begin{array}{ccr}
    \begin{tabular}{m{4cm} m{4cm} m{4cm}}
        \toprule
        Orbital Parameter & Initial Value & Value after 12 h due to $J_{2}$\\
        \midrule
        $a$(km) &  $14,900$ &   $14,900$   \\
        	$i(^\circ)$ & $67$  &   $67$   \\
        $e$	& $0.5$  &   $0.5$ \\
        $\Omega(^{\circ})$	& $46$  &   $45.8668$ \\
        $\omega(^{\circ})$ & $70$ &  $69.9597$ \\
        $M(^{\circ})$ & $330.373$ &   $1189.5$
    \end{tabular}
    \end{array}$
    \end{center}
    \caption{Variation of orbital elements in \cite{Fromm2021} due to $J_{2}$ effect}
        \label{tab:orbj2fromm}
\end{table}
\\We next investigate the proposal of \citet{Rudnit:2023}. The authors considered deployment of two or more orbiters in near-Earth circular orbits so as to rapidly fill up the $(u,v)$ plane. In echoing the results of \citet{Kudriashov:2021}, it was stated that the choice of the Right Ascension of Ascending Node (RAAN) of $-43^{\circ}$ can provide similar $(u,v)$ coverage for \m{} and \s{}. Now, we know that RAAN (which is $\Omega$ in our notation) is indeed affected by the $J_{2}$ perturbation and based on the reasoning discussed earlier, the effect is expected to be much more prominent for near-Earth orbits. The variation of specifically the RAAN is shown in Table \ref{tab:four}.
\begin{table}[h] 
\begin{center}
$\begin{array}{ccccc}
    \begin{tabular}{m{3cm} m{3cm} m{3cm} m{3cm} m{4cm}}
        \toprule
        Orb. Param. & Initial Value & After 12h & After 24h & After 12 days  \\
        \midrule
        $a$(km)	& $7500$  &   $7500$ & $7500$ & $7500$ \\
        $e$	& $0$  &   $0$ & $0$ & $0$ \\
        $i(^{\circ})$	& $-61$  &   $-61$ & $-61$ & $-61$ \\
        $\Omega_{1}(^{\circ})$ & $-43$ & $-44.369$ & $-45.739$ & $-75.882$\\
        $\Omega_{2}(^{\circ})$ & -82.29 & -83.659 & -85.0297 & $-115.173$
    \end{tabular}
    \end{array}$
    \end{center}
    \caption{Variation of orbital elements in \citet{Rudnit:2023} due to $J_{2}$ effect. Here $\Omega_{1}$ is the RAAN identified as giving similar $(u,v)$ coverage for \m{} and \s{}, while $\Omega_{2}$ is to make \s{} the primary target.}
    \label{tab:four}
\end{table}
It is therefore quite evident that the $J_{2}$ perturbation has a non-trivial effect on the RAAN for LEO deployments. Since the choice of RAAN was to observe a specific astronomical source (both \m{} and \s{} for $\Omega_{1}$ or just \m{} for $\Omega_{2}$), if the $J_{2}$ effect is not taken into account, there will be significant, unexpected variation in the visibility of target sources. This would effect the times at which the source is in view of the interferometer and the achievable variation in baseline length. 

Lastly, we consider the proposal of \citet{Andrianov:2021} that considered the orbiter to be on a Highly Elliptical Orbit (HEO).
\begin{table}[h]
\begin{center}
    $\begin{array}{ccc}
    \begin{tabular}{m{4cm} m{4cm} m{4cm}}
        \toprule
        Orb. Param. & Initial Value & After 10 days (T1;$J_{2}$)   \\
        \midrule
        $a$(km)	& $165000$  &   $165000$ \\
        $e$	& $0.939$  &   $0.939$ \\
        $i(^{\circ})$ & $20.08$  &   $20.08$ \\
        $\Omega(^{\circ})$ &  $-3.583$ &   $-3.59198$   \\
        	$\omega(^{\circ}$) & $-92$  &   $-91.9837$
    \end{tabular}
    \end{array}$
    \end{center}
    \caption{Variation of orbital elements in Orbit Type 1 of \citet{Andrianov:2021} due to $J_{2}$ effect}
    \label{tab:orbj2andrian}
\end{table}
Since the orbiter is in a HEO, the effect of $J_{2}$ is less prominent here. Nevertheless, the fact that the proposal requires highly elliptical orbits, mitigation of the $J_{2}$ effect through orbit adjustment would lead to a non-trivial contribution to the fuel budget of the orbiter \citep{Lee:2022}. A notable feature of the study in \citet{Andrianov:2021} is that they use the EGM96 Earth gravity model which does indeed take into account the non-central nature of the Earth's gravitational field (including of course, the $J_{2}$ effect). This is similar to the Earth-space VLBI mission with RadioAstron being the space-based component \citet{Kardashev:2013}, which also took into account the non-central gravitational field of the Earth, along with perturbations due to the Moon and the Sun, noting that the presence of such factors ``substantially complicated determination of the spacecraft orbit"\citep{Kardashev:2013}.
\\
From investigating the tables, several general inferences can be obtained. Firstly, the fact that the $J_{2}$ effect seems ``negligible" in some cases is a consequence of the fact that the proposals discussed here considered orbiters in and around the Medium Earth Orbit (MEO), wherein the $J_{2}$ effect is smaller when compared to the LEO case. This is straightforward to infer from a physical sense because the farther the orbiter goes from the Earth, the lesser will be the effect of Earth's gravitational field on it. Nevertheless, the $J_{2}$ effect does indeed contribute to the change in values of the orbital parameters which in several cases have been obtained from optimisation algorithms catered to observing a particular source, for example \s{} by \citet{Fromm2021}. The fact that the $J_{2}$ effects reduce as the altitude of the orbiter increases is apparent. Therefore, since the primary impetus of going to larger baselines (and therefore to MEO and beyond) was to obtain sharpened angular resolution, it is important to investigate how specific perturbing effects can influence these science goals. As far as LEO deployments are concerned, the importance of mitigating the $J_{2}$ perturbation is quite evident from Table \ref{tab:four}. \\
The lessons drawn here are analogous to ones obtained from orbit determination studies for RadioAstron \citep{Zakhvatkin:2020}. In addition to the evolution of the orbital parameters due to Earth's gravitational field, it was noted that the solar radiation pressure was the main perturbing force affecting the orbiter's motion. The modelling of the orbiter's dynamics specifically taking this effect into account was crucial to studying the orbiter's efficacy for conducting observations during the mission's observing cycle. Therefore, one can envision that the $J_{2}$ effect can play a similar role in selecting orbits for LEO deployments of space VLBI missions.

\subsection[The J2 effect and photon ring observations]{The $J_{2}$ effect and photon ring observations}
\begin{figure}
    \centering
    \includegraphics[width=0.7\columnwidth]{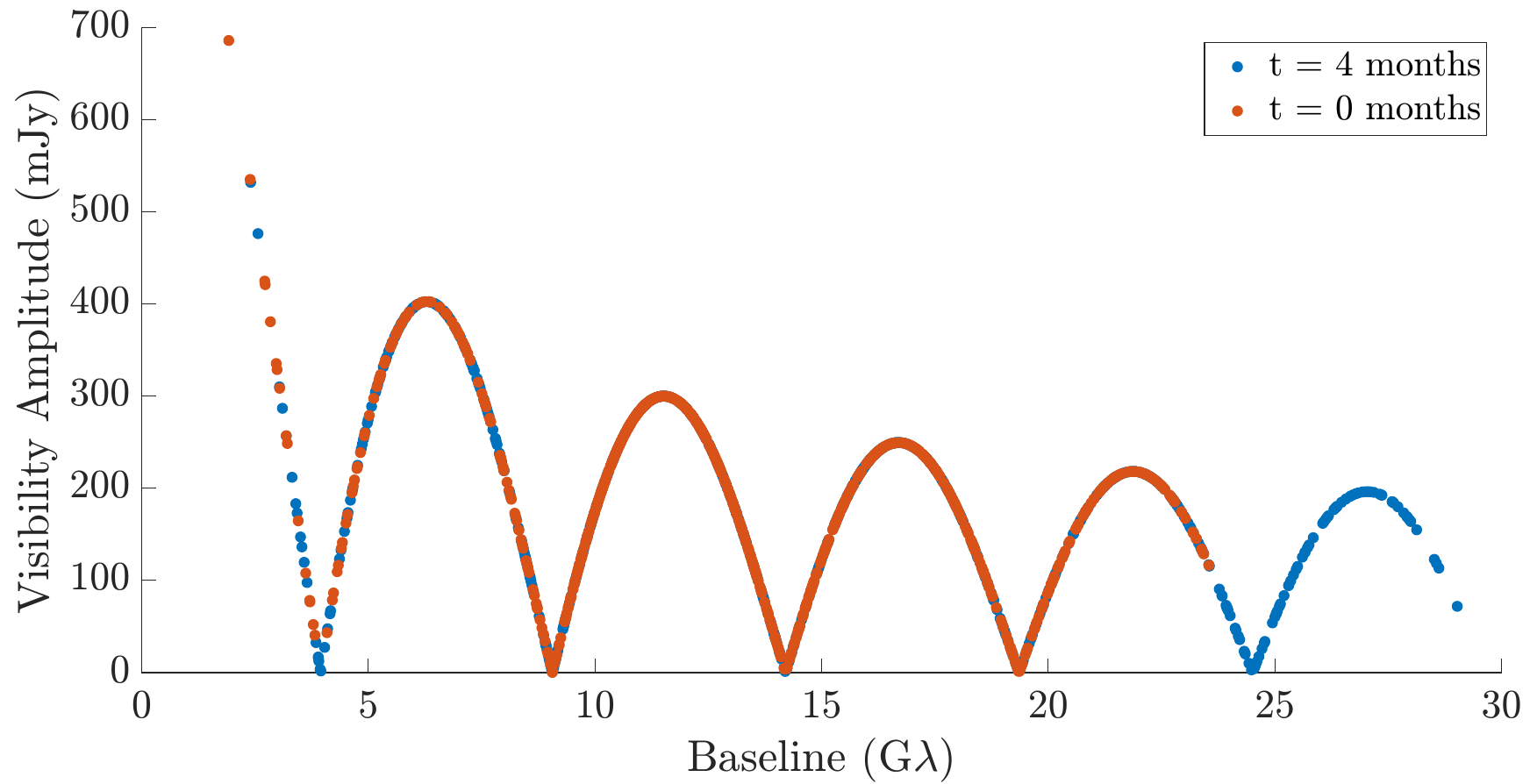}
    \caption{Interferometric signature of an infinitesimally-thin, circular ring, using parameters consistent with GRMHD simulations of \m{}, demonstrating the variation in coverage caused by $J_{2}$-induced orbit precession. Observations conducted over 15 days with same duty cycle and integration time parameters as used for Fig. \ref{f:SunSynchUV}, Fig. \ref{f:MolniyaUV} and \ref{f:EquatorialUV}.}
    \label{f:visAmpVariation}
\end{figure}

The high resolution observations of a black hole photon ring is one of the prime targets for future Earth-space and space-only VLBI observations \citep{johnson_universal_2020,Fish2020,Lupsasca:2020,Peter:2022, Hudson_2023}. Indeed, one of the primary motivations for putting an orbiter into space for VLBI observations is to obtain baselines which are much longer than the ones possible using solely Earth-based stations. Keeping this in mind, it is imperative to investigate whether the $J_{2}$ effect impacts the observations of the photon ring. \\
To obtain a handle on these effects, we attempt to model the changes in the $(u,v)$ plane due to the presence of the $J_{2}$ effect directly into photon ring observations in the visibility domain. In this domain, the interferometric signature $V(u)$ of an infinitesimally thin, uniform circular ring for a  diameter $d$ (measured in radians), observed on $(u,v)$-distances $u$ (measured in wavelengths), is given by the zeroth order Bessel function of the first kind:
\begin{gather}
    V=J_{0}(\pi d u).
\end{gather}
This ring has a unit total flux. Fig. \ref{f:visAmpVariation} depicts the variation in coverage of a model photon ring about \m{} due to the $J_{2}$ perturbation on the equatorial HEO interferometer configuration.
On an Equatorial HEO, the inclusion of the $J_{2}$ effect leads to a preferable increase in the maximum baseline length achieved, after only 4 months of precession. The density of coverage over the shorter baselines is however reduced. The precession of different orbit configurations would no doubt result in a more drastic variation in the baseline coverage of the interferometer over time. The equatorial HEO example does however show that an appreciable difference in the interferometric signature can be observed due to the precession of the orbit. Therefore, this must be considered when planning observations of such a mission as depending on the current rotation of the right ascension about the equator, the baseline range achievable will vary.


\section{Incorporating equations of motion of orbiters into the \texorpdfstring{\lowercase{$(u,v)$} space}{(u,v)}}\label{sec4}
In radio interferometry and therefore in VLBI as well, the standard method to judge the efficacy of a baseline for various scientific goals is to investigate the coverage in the so-called $(u,v)$ space \citep{Thom}. Since the astrodynamics effects, in particular the Earth's oblateness being discussed here, directly impact the equations of motion of the orbiter, one requires a set of equations that directly map these equations to the coverage in the $(u,v)$ plane. \\
The astrodynamic considerations become all the more important for space-only VLBI observations since the astrodynamical effects would now affect \textit{each and every} orbiter in a potentially distinct manner (based on the orbiter design, orbital parameters etc.), as compared to a simple Earth-Space VLBI deployment of a single orbiter. Therefore, it is even more important for space VLBI to have a mapping that allows for a suitable translation between astrodynamical effects and the associated consequences on the $(u,v)$ plane. \\
For this paper, we shall consider the case of two orbiters in relative motion with one another, thereby forming a space-only baseline. In borrowing nomenclature from astrodynamics \citepalias{Sengupta}, one orbiter is labelled as the Chief and the other orbiter is the Deputy. The equations of motion are modelled for the motion of the Deputy relative to that of the Chief. The analysis proceeds as follows.\\
We require two main co-ordinate systems, namely the ECI frame discussed earlier and the Local Vertical Local Horizontal (LVLH) frame which has the origin at the orbiting satellite \citepalias{Sengupta}.
Now, if $(H,\delta)$ are the hour-angle and declination respectively of the astrophysical source under observation, and $\lambda$ is the observing wavelength, the $(u,v,w)$ co-ordinates used in interferometry can be obtained from the components ($X,Y,Z$) components in the ECI frame by:
\begin{gather}
    \begin{pmatrix}
u\\v\\w
    \end{pmatrix}=\frac{1}{\lambda}\begin{pmatrix}
        \sin H&\cos H&0\\-\sin \delta\cos H &\sin\delta\sin H &\cos\delta\\\cos\delta\cos H &-\cos\delta\sin H & \sin\delta
    \end{pmatrix}\begin{pmatrix}
        X\\Y\\Z
    \end{pmatrix}.
\end{gather}
For applications in VLBI, we only require the $(u,v)$ co-ordinates since the contributions from the $w$ co-ordinate are considered negligible, which arises from the assumption that the field of the source being synthesised is not too large (see Chapter 3 of \citet{Thom} for an extended discussion on this matter). Therefore, we shall focus only on $(u,v)$ and the corresponding equations are:
\begin{gather}
    u=\frac{1}{\lambda}(X\sin H +Y\cos H),\\ \nonumber
    v= \frac{1}{\lambda}(-X\sin\delta\cos H +Y\sin\delta\sin H +Z\cos\delta). \label{eq:uv}
\end{gather}
Now, let $\vec{r}_{c}$ be the position vector components of the chief in the ECI system:
\begin{gather}
    \vec{r}_{c}=X_{c}\hat{X}+Y_{c}\hat{Y}+Z_{c}\hat{Z},
\end{gather}
and let the corresponding components in the LVLH frame be $(X_{c0},Y_{c0},Z_{c0})$.Three of its orbital elements $\Omega,i$ and $\theta$ form a $3-1-3$ Euler system ($\Omega-i-\theta)$  and so the aforementioned components of its position vector in the ECI frame can be written in terms of the LVLH components using a directional cosine matrix \citepalias{Sengupta}:
\begin{gather}
\begin{pmatrix}
    X_{c}\\Y_{c}\\Z_{c}
\end{pmatrix}=\begin{pmatrix}
    \cos\theta_{c}\cos\Omega_{c}-\sin\theta_{c}\cos i_{c}\sin\Omega_{c}&-\sin\theta_{c}\cos\Omega_{c}-\cos\theta_{c}\cos i_{c}\sin\Omega_{c}&\sin i_{c}\sin\Omega_{c}\\ \cos\theta_{c}\sin\Omega_{c}+\sin\theta_{c}\cos i_{c}\cos\Omega_{c}&-\sin\theta_{c}\sin\Omega_{c}+\cos\theta_{c}\cos i_{c}\cos\Omega_{c}&-\sin i_{c}\sin\Omega_{c}\\ \sin\theta_{c}\sin i_{c} &\cos\theta_{c}\cos i_{c} & \cos i_{c}
\end{pmatrix}\begin{pmatrix}
    X_{c0}\\Y_{c0}\\Z_{c0}
\end{pmatrix}.
\end{gather}
Here $\theta_{c},i_{c},\Omega_{c}$ are the latitude, inclination and right ascension respectively for the chief. Expanding the equation, we get:
\begin{gather}
    X_{c}=(\cos\theta_{c}\cos\Omega_{c}-\sin\theta_{c}\cos i_{c}\sin\Omega_{c})X_{c0}+(-\sin\theta_{c}\cos\Omega_{c}-\cos\theta_{c}\cos i_{c}\sin\Omega_{c})Y_{c0}+(\sin i_{c}\sin\Omega_{c})Z_{c0},\nonumber\\ 
    Y_{c}=(\cos\theta_{c}\sin\Omega_{c}+\sin\theta_{c}\cos i_{c}\cos\Omega_{c})X_{c0}+(-\sin\theta_{c}\sin\Omega_{c}+\cos\theta_{c}\cos i_{c}\cos\Omega_{c})Y_{c0}+(-\sin i_{c} \sin\Omega_{c})Z_{c0},\nonumber \\ 
    Z_{c}=(\sin\theta_{c}\sin i_{c})X_{c0}+(\cos\theta_{c}\cos i_{c})Y_{c0}+(\cos i_{c})Z_{c0}. \label{eq:complvlh}
\end{gather}
If we now define $(X_{d},Y_{d},Z_{d})$ as the position vector components of the deputy in the ECI frame, we can define its $(u,v)$ co-ordinates as in Equation \ref{eq:uv}. Then, using an analogous expression for the chief, we can define
\begin{gather}
    \label{eq:uvdc}
    u\equiv u_{d}-u_{c}=\frac{1}{\lambda}((X_{d}-X_{c})\sin H+(Y_{d}-Y_{c})\cos H),\\ \nonumber
    v\equiv v_{d}-v_{c}=\frac{1}{\lambda}(-(X_{d}-X_{c})\sin\delta\cos H+(Y_{d}-Y_{c})\sin\delta\sin H+(Z_{d}-Z_{c})\cos\delta).
\end{gather}
We now write this equation in the LVLH frame. In this frame, suppose the components of the separation vector between the chief and deputy are given by:
\begin{gather}
    \label{eq:diffvec}
    X_{d0}-X_{c0}=x,\quad Y_{d0}-Y_{c0}=y,\quad Z_{d0}-Z_{c0}=z .
\end{gather}
Then, using Equation \ref{eq:complvlh} for the chief, and exactly the same construction for the deputy (since both are in the LVLH frame, and $\Omega,i,\theta$ transform any general vector from the ECI to the LVLH frame, the angles would be the same for both the chief and the deputy), substituting in Equation \ref{eq:uvdc} along with Equation \ref{eq:diffvec} for the difference in the position vector components, we get our main equations: 
\begin{gather}
    u(\lambda)=\frac{1}{\lambda}\Biggl\{\sin H\Bigg[(cos\theta_{c}\cos\Omega_{c}-\sin\theta_{c}\cos i_{c}\sin\Omega_{c})x+(-\sin\theta_{c}\cos\Omega_{c}-\cos\theta_{c}\cos i_{c}\sin\Omega_{c})y\nonumber\\
    +(\sin i_{c}\sin\Omega_{c})z\Bigg]+\cos H\Bigg[(\cos\theta_{c}\sin\Omega_{c}+\sin\theta_{c}\cos i_{c}\cos\Omega_{c})x+(-\sin\theta_{c}\sin\Omega_{c}+\cos\theta_{c}\cos i_{c}\cos\Omega_{c})y \nonumber \\-(\sin i_{c}\sin\Omega_{c})z\Bigg]\Biggr\},\nonumber \\
    v(\lambda)=\frac{1}{\lambda}\Biggl\{-\sin\delta\cos H\Bigg[(cos\theta_{c}\cos\Omega_{c}-\sin\theta_{c}\cos i_{c}\sin\Omega_{c})x+(-\sin\theta_{c}\cos\Omega_{c}-\cos\theta_{c}\cos i_{c}\sin\Omega_{c})y\nonumber \\
    +(\sin i\sin\Omega_{c})z\Bigg]+\cos\delta\Bigg[(\cos\theta_{c}\sin\Omega_{c}+\sin\theta_{c}\cos i_{c}\cos\Omega_{c})x+(-\sin\theta_{c}\sin\Omega_{c}+\cos\theta_{c}\cos i_{c}\cos\Omega_{c})y\nonumber \\ -(\sin i_{c}\sin\Omega_{c})z\Bigg]
    +\cos\delta\Bigg[(\sin\theta_{c}\sin i_{c} )x+(\cos\theta_{c}\cos i_{c})y+(\cos i_{c})z\Bigg]\Biggr\}\label{eq:uvbase}.
\end{gather}
The Equations \ref{eq:uvbase}, explicitly derived to the best of our knowledge for the first time, will be the main tools used to incorporate astrodynamics effects into space-only VLBI. In particular, one notices that due to the mapping between the equation of motion of an orbiter, given by $(x,y,z)$, and the $(u,v)$ plane, the realistic effects incorporated in the former can now be reflected in the latter.\\
Lastly, one can note that since the origin of the chief's LVLH frame has $(x=y=z=0)$, the ($u,v$) co-ordinates are $(0,0)$. Thus, the  $(u,v)$ plane is ``centered" on the chief orbiter's LVLH frame.

\section[Equations for Relative Motion for Chief and Deputy: No J2 effect]{Equations for Relative Motion for Chief and Deputy: No $J_{2}$ effect}\label{sec5}
Let $\rho$ be the relative orbit radius. Then, under the assumption of having a circular reference orbit and no perturbing forces, the relative equations of motion are given by the Hill-Clohessey-Wiltshire (HCW) Equations \citep{Schaub}:
\begin{gather}
    \ddot{x}-2n_{c}\dot{y}-3n_{c}^{2}x=0,\nonumber \\
    \ddot{y}+2n_{c}\dot{x}=0, \nonumber \\
    \ddot{z}+n_{c}^{2}z=0
\end{gather}
where $n_{C}$ is the mean motion for the chief obtained by substituting the orbital parameter $r_{C}$ of the chief into $r$ in \ref{eq:meanmotion}. Notice that the $(x,y)$ motion is decoupled from the $z-$motion. The former can be modelled as coupled harmonic oscillator while the latter can be modelled as a harmonic oscillator. \\
For bounded motion with relative orbit radius $\rho$ and initial conditions such that the orbits trace a circle in the $(y-z)$ plane, and assuming no offset in the  the $x$ and $y$ direction, solution to these equations is given by:
\begin{gather}
    x(t)=\rho\cos (n_{c}t +\alpha), \nonumber\\
    y(t)=-2\rho\sin(n_{c}t+\alpha), \nonumber\\
    z(t)=2\rho\cos(n_{c}t+\alpha). \label{eq:HCWeom}
\end{gather}
\subsection{ \texorpdfstring{\lowercase{$(u,v)$} plots}{(u,v)}}
\begin{figure}[h] 
    \centering
    \includegraphics[width=0.5\textwidth]{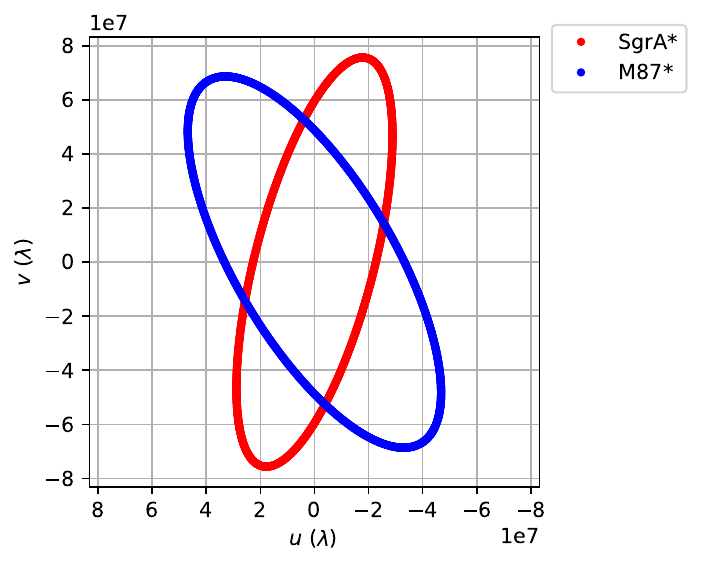}
    \caption{$(u,v)$ coverage of \m{} and \s{} at $f=345$ GHz using HCW equations for a time period of 6 months.}
    \label{fig:uvhcw}
\end{figure}

We now construct the $(u,v)$ plots using the equations of motion given in Equation \ref{eq:HCWeom}. The choice of parameters are given in Table \ref{tab:hcwparam}. The numerical values are inspired from the ones given in \citetalias{Sengupta} and the form of the solutions is obtained using the equations in \citet{Ginn}.
\begin{table} 
\begin{center}
$\begin{array}{cc}
    \begin{tabular}{m{2cm} m{2cm} }
        \toprule
        Orb. Param. & Value   \\
        \midrule
        $a_{c}$(km)	& $7000$  \\
        $e_{c}$	& $0$   \\
        $i_{c}(^\circ)$ & $35$ \\
        $\theta_{c}(^\circ)$	& $0$   \\
        $\Omega_{c}(^\circ)$ &  $0$    \\
        \bottomrule
    \end{tabular}
    \end{array}$
    \end{center}
    \caption{Orbital Parameters of the Chief}
    \label{tab:hcwparam}
\end{table}
In addition, we consider the mean relative orbital radius $\rho$ and the offset $\alpha$ to be
\begin{gather}
    \rho=25\text{km},\quad \alpha=0.
\end{gather}
Lastly, we consider observations to be at $345$ GHz.
Substituting all of these parameters into the Equations \ref{eq:uvbase}, we obtain the $(u,v)$ tracks shown in Figure \ref{fig:uvhcw}.
It can be noticed that the baselines obtained through this formalism are not large enough for high angular resolution observations required for say resolving the photon ring \citep{johnson_universal_2020}. This should be expected because the HCW equations were obtained using a linearisation argument because of which the order of separation between the chief and deputy ($~10$'s of km) has to be much less than the semi-major axis of the chief ($~1000$'s of km). Therefore, while changing the values of $\rho$ to higher values by ``brute force" will give ``bigger" ellipses in the $(u,v)$ plane, it would not be faithful to the physical basis of the HCW solutions. Nevertheless, one can still obtain such elliptical plots in the $(u,v)$ plane using a modified set of equations that make the orbits invariant to certain effects of the $J_{2}$ perturbation. This will be discussed in later sections.

\section[Equations for Relative Motion for Chief and Deputy: Incorporating J2 Corrections through Linearisation]{Equations for Relative Motion for Chief and Deputy: Incorporating $J_{2}$ Corrections through Linearisation} \label{sec6}
We now discuss how the effects of $J_{2}$ perturbation can be incorporated using linearisation of the equations of motion. The formalism for the same has been provided in \citet{Schweig,Ginn} and it has been provided in the Appendix. In the subsequent sections, we discuss the applications of the formalism to obtain $(u,v)$ plots of \m{} and \s{}.\\

\subsection{ \texorpdfstring{\lowercase{$(u,v)$} plots}{(u,v)}}
\begin{figure*}
    \centering
    \includegraphics[width=\textwidth]{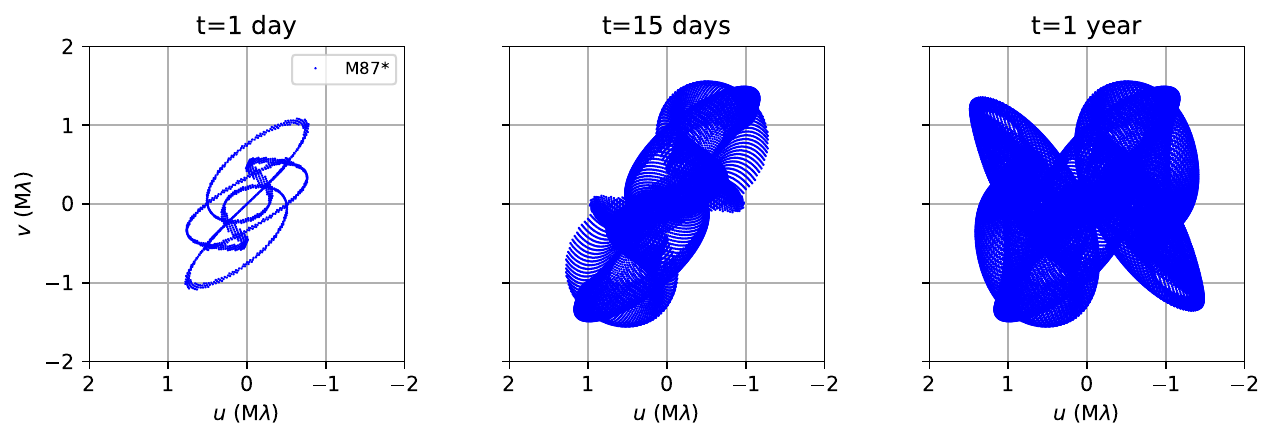}
    \caption{$(u,v)$ plots for \m{} $J_{2}$ effects and cross-track drift corrections. The time periods are, from left to right, 1 day, 15 days and 1 year.}
    \label{fig:uvj2ctcm87}
\end{figure*}

\begin{figure*}
    \centering
    \includegraphics[width=\textwidth]{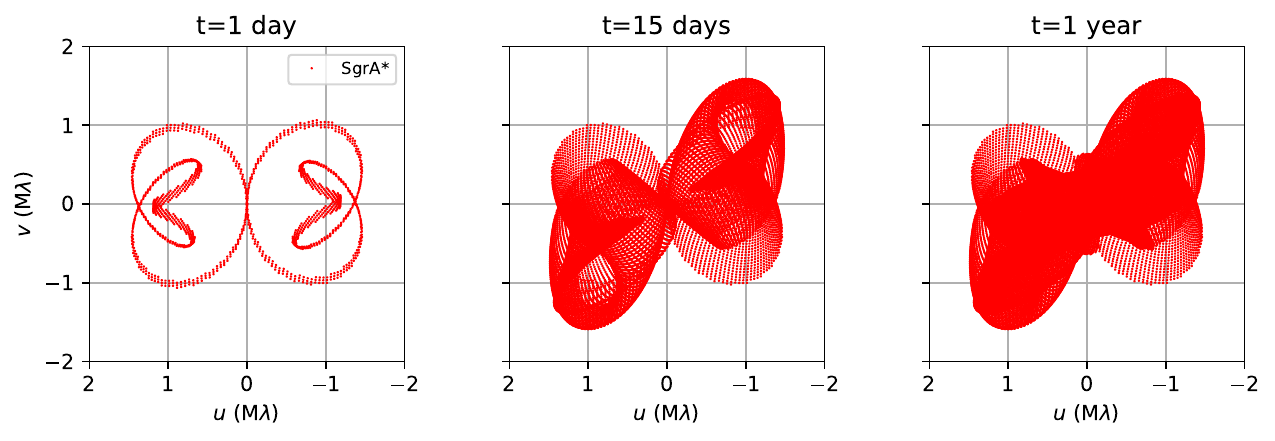}
    \caption{$(u,v)$ plots for \s{} including $J_{2}$ effects and cross-track drift corrections. The time periods are, from left to right, 1 day, 15 days and 1 year.}
    \label{fig:uvj2ctsgra}
\end{figure*}

The $(u,v)$ plots \m{} are given in Figure \ref{fig:uvj2ctcm87} whilst for \s{} are given in Figure \ref{fig:uvj2ctsgra}. Several inferences can be drawn.


 Firstly, the plots show a remarkably rich and distinctive character based on the black hole that is being observed, namely \m{} and \s{}. In particular, these plots follow directly from the equations of motion without use of any ``optimisation" procedures, thereby allowing us to directly capture the impact of the $J_{2}$ effect on the $(u,v)$ coverage. Secondly, as time passes, the plots become increasingly dense and thus provide dense $(u,v)$ on short baselines, a desired feature of space-VLBI missions \citep{Fish2020}. We note that in order to generate these plots, only a small sample of the data points have been used that were generated by the choice of the sampling rate. For time periods of 1 day, 15 days and 1 year, we used 700, 15,000 and 30,000 data points respectively. This choice was governed largely by requiring a reasonable processing time to generate and process the plots in a higher resolution. It has been verified that utilisation of all the plot points leads to a modest increase in filling  $(u,v)$ in the ``North-West" and ``South-East" parts of the plot for \m{} and ``North-South" parts for \s{}. In addition, there is an overall increase in the coverage density. Nevertheless, the existing plots still demonstrate the key scientific takeaway of rapidly increasing density over medium-length baselines providing a rich geometric coverage pattern generated primarily by the $J_{2}$ effect \\
The extremely dense coverage over shorter baselines introduces the possibility of having an ``ALMA-like" flying formation of orbiters in space that can have much longer integration times and observe every single day throughout the year without having to mitigate intra-day variability in weather. As another potential avenue, one can consider a ``hybrid" mission wherein  one one hand, the chief and/or the deputy can work in an ``Earth-Space" mode, forming long baselines allowing high angular resolution images of the photon ring, as discussed in the earlier sections on Earth-space VLBI. On the other hand, the the orbiters can operate in a ``Space-only" more wherein the relative baselines formed between them as part of a formation can allow for observations of large-scale structure pertaining to black hole accretion. While providing a detailed description of a mission design based on these principles is beyond the scope of this paper, the results here indicate that in both of these considerations, the $J_{2}$ effect can turn out to be advantageous in not only providing dense $(u,v)$ coverage, but also aiding in suitable orbit selection. 
\section[Equations for Relative Motion for Chief and Deputy: J2 Invariant Orbits]{Equations for Relative Motion for Chief and Deputy: $J_{2}$ Invariant Orbits} \label{sec7}
There have been several investigations into obtaining suitable orbital parameters such that the motion of the orbiter is unaffected by the $J_{2}$ perturbation \citep{Schaub:2001,Lee:2022}. In this section, we shall largely be following the discussion given in \citet{Lee:2022}.\\
The $J_{2}$ perturbation $independently$ affects the latitude $\theta$ and the RAAN $\Omega$ and so finding $J_{2}$ invariant orbits boils down to searching for orbital parameters that can minimise their secular change over time, as given in Equations \ref{eq:Omj2}, \ref{eq:omj2} and \ref{eq:Mj2}. Suppose we consider the motion of the chief. If one has to minimise the effect of the $J_{2}$ perturbation on the chief, then the following equation must be solved
\begin{gather}
    \dot{\theta}_{c}=\dot{\omega}_{c}+\dot{M}_{c}=0. \label{eq:minlatj2}
\end{gather}
Now, for the subsequent discussion, we shall consider the chief to be in a circular orbit, and so
\begin{gather}
    e_{c}=0 \label{eq:chiefecc}.
\end{gather}
Upon substituting the expressions from \ref{eq:omj2} and \ref{eq:Mj2} into \ref{eq:minlatj2}, and using the circular orbit condition \ref{eq:chiefecc} we get \citep{Lee:2022}:
\begin{gather}
   \dot{\theta}= \frac{3}{2}J_{2}n_{c}\Bigg(\frac{R_{e}}{a_{c}}\Bigg)^{2}(4\cos^{2}i_{c}-1)=0.
\end{gather}
Since the terms outside the bracket are constants, this equation can only be satisfied when the term in the bracket is $0$:
\begin{gather}
    4\cos^{2}i_{c}-1=0\Rightarrow i_{c}= \pm60^{\circ}.
\end{gather}
Thus, if one wishes to minimise the change in latitude for the chief in a circular orbit, that can be achieved by having it at an inclination angle of $60^{\circ}$. It is important to note that this result holds for \textit{all} altitudes of the chief satellite. Note that this is analogous to our earlier calculation in Equation \ref{eq:molniyamimj2} on having an inclination of $63.4^{\circ}$ that minimised $\dot{\omega}$ for the Molniya orbit.\\
Now, if the change in latitude due to the $J_{2}$ perturbation has to be minimised, we have:
\begin{gather}
    \dot{\theta}_{d}=\dot{M}_{d}+\dot{\omega}_{d}=0.\nonumber\\
\end{gather}
Using once again the Equations \ref{eq:omj2} and \ref{eq:Mj2}, we get \citep{Lee:2022}:
\begin{gather}
    \dot{\theta}_{d}=\frac{3}{4}J_{2}n_{d}\Bigg(\frac{R_{e}}{p_{d}}\Bigg)^{2}\Bigg[(5+3\sqrt{1-e_{d}^{2}})\cos^{2}i_{d}-(\sqrt{1-e_{d}^{2}}+1)\Bigg]=0.
\end{gather}
Solving for $i_{d}$, we have the condition \citep{Lee:2022}:
\begin{gather}
    i_{d}=\pm\cos^{-1}\Bigg(\sqrt{\frac{1+\sqrt{1-e_{d}^{2}}}{5+3\sqrt{1-e_{d}^{2}}}}\Bigg) \label{eq:deputyminj2}.
\end{gather}
Thus for a given value of the eccentricity of the deputy, the inclination that minimises the $J_{2}$ perturbation can be obtained by solving the Equation \ref{eq:deputyminj2}. One can also do similar calculations for minimising the RAAN, which as we mentioned earlier has an independent effect on the equations of motion, but in this paper we focus on the effects of minimising the change in latitude, postponing the RAAN based discussion to a future work. \\
Now, minimising the drift in $\theta$ has a direct impact on the fuel budget of the mission. Under an impulsive control scheme that aims to maintain constant relative motion between the deputy and chief during the mission lifetime, the so-called $\Delta V$ requirement for the scheme for latitude correction, exercised after one orbit, is given by \citep{Lee:2022}:
\begin{gather}
    \Delta V_{\dot{\theta}} =\frac{a_{c}}{3}\Big[(\dot{\omega}_{d}-\dot{\omega}_{c})+(\dot{M}_{d}-\dot{M_{c}})\Big].
\end{gather}
Thus, if both $\dot{M}_{c}+\dot{\omega}_{c}=0$ and $\dot{M}_{d}+\dot{\omega}_{d}=0$ hold, we have
\begin{gather}
    \Delta V_{\dot{\theta}}=0,
\end{gather}
and so additional fuel will not be dispensed to maintain relative motion. Similar considerations apply for minimising the RAAN \citep{Lee:2022}.

\subsection{ \texorpdfstring{\lowercase{$(u,v)$} plots for $J_{2}$ invariant orbits}{(u,v)}}
We now obtain the $(u,v)$ plots for $J_{2}$ invariant relative motion using suitable parameter substitution in Equation \ref{eq:uvbase}.\\
Firstly, for the chief we consider the parameters given in \ref{tab:orbchiefj2inv}. The mean anomaly $M_{c}$ has not been mentioned since it does not play a role in the subsequent discussions.
\begin{table}[h] 
\begin{center}
$\begin{array}{ccccc}
    \begin{tabular}{m{2cm} m{2cm} m{2cm} m{2cm} m{2cm}}
        \toprule
        $a_{c}(km)$ & $e_{c}$ & $i_{c}^{\circ}$ & $\theta_{c0}^{\circ}$ & $\Omega_{c0}^{\circ}$   \\
        \midrule
        $7000$	& $0$  &   $60$ & 0 & 0 
    \end{tabular}
\end{array}$
\end{center}
\caption{Orbital Parameters of the Chief for $J_{2}$ Invariant Orbits}
 \label{tab:orbchiefj2inv}
\end{table}
The choice of $i_{c}=60^{\circ}$ implies that there would be no change in the latitude $\theta_{c}$ over time. We thus assume:
\begin{gather}
    \theta_{c}(t)\equiv\theta_{c0}, \quad \Omega_{c}(t)\equiv\Omega_{c0}.
\end{gather}

Since the latitude and RAAN drift independently affect the relative motion under the $J_{2}$ effect, one can only design a configuration that minimises either one of them. Indeed, the equations to solve for inclinations that minimise the $J_{2}$ effect on RAAN are different than those for minimising the effect in the latitude \citep{Lee:2022}. \\
Lastly, we do allow for mean anomaly of the deputy to have a time dependence as per the Equation \ref{eq:varorb}.\\
Now, following \citet{Lee:2022}, the equation of motion of the deputy in the chief's frame, for the case that both chief and deputy have the same altitude, with the chief having a small eccentricity, modelled so that there is zero relative latitude drift, is:
\begin{gather}
x(t)=-a_{d}\cos\Bigg[M_{d}(t)\sqrt{1-\Bigg(\frac{1-5\cos^{2}i_{d}}{3\cos^{2}i_{d}-1}\Bigg)^{2}}\Bigg], \nonumber \\
y(t)=2a_{d}\sin\Bigg[M_{d}(t)\sqrt{1-\Bigg(\frac{1-5\cos^{2}i_{d}}{3\cos^{2}i_{d}-1}\Bigg)^{2}}\Bigg]. \label{eq:depeom}
\end{gather}
wherein $M_{d}(t)$ is governed by the Equation \ref{eq:varorb}. For the conditions discussed here, the motion along $z$ direction is negligible \citep{Lee:2022} and so we assume
\begin{gather}
    z(t)=0.
\end{gather}
We now describe in detail the choices for parameters of the deputy. Firstly, since the chief and deputy have to have the same altitude and hence the same semi-major axis, we have:
\begin{gather}
    a_{d}=a_{c}=7000km.
\end{gather}
Next, we assume the deputy to have a small eccentricity of
\begin{gather}
    e_{d}=0.08,
\end{gather}
and using Equation \ref{eq:deputyminj2}, we get the value for inclination to be (taking only the positive value):
\begin{gather}
    i_{d}=60.0066^{\circ}.
\end{gather}
We choose the initial value of the mean anomaly to be
\begin{gather}
    M_{d0}=0^{\circ},
\end{gather}
so that at a later time $t$, we have:
\begin{gather}
    M_{d}(t)=\Bigg(n_{d}\Bigg[1+\frac{3}{4}\sqrt{1-e_{d}^{2}}J_{2}\Bigg(\frac{R_{e}}{p_{d}}\Bigg)^{2}(3\cos^{2}i_{d}-1)\Bigg]\Bigg)t+M_{d0}, \nonumber\\
    =\Bigg(n_{d}\Bigg[1+\frac{3}{4}\sqrt{1-e_{d}^{2}}J_{2}\Bigg(\frac{R_{e}}{p_{d}}\Bigg)^{2}(3\cos^{2}i_{d}-1)\Bigg]\Bigg)t.
\end{gather}
Upon substituting the numerical values for the parameters, we get:
\begin{gather}
    M_{d}^{\circ}(t)=0.0618367^{\circ}t. \label{eq:mdt}
\end{gather}
Then, using Equation \ref{eq:mdt} and substituting the value for $i_{d}$ into Equation \ref{eq:depeom}, we get the final equations of motion:
\begin{gather}
    x(t)=-7000\cos(0.0618367^{\circ}t\times 0.08)=-7000\cos(0.00494693^{\circ}t),\nonumber\\
    y(t)=14,000\sin(0.00494693^{\circ}t). \label{eq:fineom}
\end{gather}
These are the final equations of motion that shall now be substituted into the $(u,v)$ Equations \ref{eq:uvbase}.

\subsection{ \texorpdfstring{\lowercase{$(u,v)$} plots}{(u,v)}}
\begin{figure}[h] 
    \centering
    \includegraphics[width=0.5\textwidth]{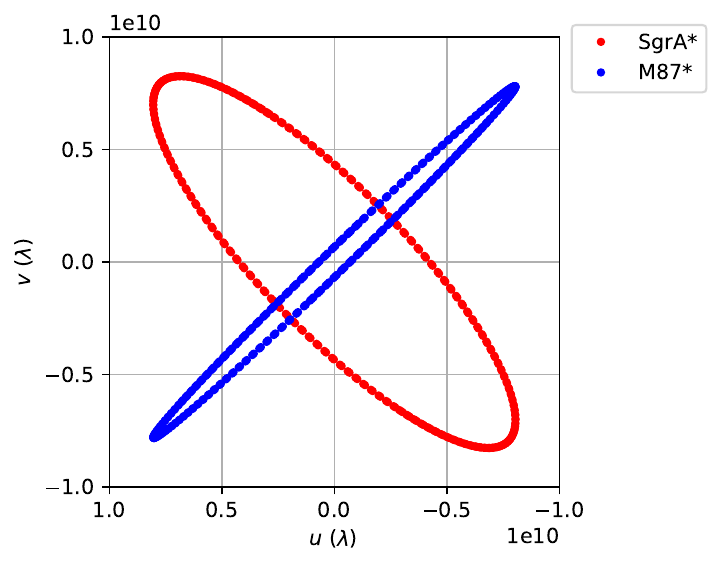}
    \caption{$(u,v)$ coverage of \m{} and \s{} using $J_{2}$ invariant equations of motion of \cite{Lee:2022} for a time period of 1 day.}
    \label{f:uvj2inv1day}
\end{figure}

The $(u,v)$ plots for \m{} and \s{} are given in Figure \ref{f:uvj2inv1day}. The observation time is chosen to be $1$ day at a sampling rate of $500s$. The choice of observing frequency is $f=345$GHz.
The plots have distinctly different elliptical shapes as compared to the ones obtained without the $J_{2}$ effect from the HCW equations in Figure \ref{fig:uvhcw}. Thus this at the very least establishes the fact that the $J_{2}$ effect in and of itself does impact the $(u,v)$ coverage of the sources under consideration. More generally, The $(u,v)$ plots look deceptively straightforward as standard VLBI plots for large baselines. However, the point we wish to convey with the plots is \textit{precisely} that, which is to say that suitable choice of parameters which are specifically aimed at minimising the $J_{2}$ effect \textit{can} give $(u,v)$ plots which are suitable for VLBI investigations. As a particular example, if a space based VLBI mission that deploys antennas in LEOs takes a simplified approach of just specifying the orbital parameters of the antenna without taking into realistic effects such as the $J_{2}$, there would a significant drift in the orbiter's motion if such effects are not taken into account, further affecting fuel budgets that would impact the mission cycle. While it is true that rigidity in the choice of orbital parameters might lead to certain compromises on the astronomical sources that can be observed, such considerations provide an impetus to investigate space-only VLBI optimisation software that does take into account realistic astrodynamical considerations. We wish to investigate these avenues in the near future.

\section{Conclusion}\label{sec8}
The paper attempts to investigate in detail the effects of the Earth's oblateness, also known as the $J_{2}$ effect, on an Earth-Space and a Space-Space VLBI mission that aims to provide black hole images of the sources \m{} and \s{}. An extensive study has been performed of several existing VLBI proposals in literature that investigated observing these sources using VLBI with a space-based component. In a similar vein, for the Earth-Space missions, a detailed discussion is provided on how an informed choice of orbital parameters based on isolating the $J_{2}$ effect can be used to one's advantage in space mission design. For the Space-Space VLBI regime, the impact of the $J_{2}$ effect is studied using a simple, computationally accessible analytic framework that studies the relative motion of a chief orbiter and a deputy orbiter in relative motion. It was found that the $J_{2}$ effect leads to distinct patterns on the $(u,v)$ plane for both \m{} and \s{} which are extremely dense over short baselines, thereby potentially opening up an avenue of providing a dense filling of the $(u,v)$ plane in space VLBI missions using just a couple of orbiters. Moreover, it was demonstrated that an informed choice of orbital parameters can also lead to bounded relative motion over longer baselines that is invariant under the $J_{2}$ perturbation. 
\subsection{Future Work}
There is a smorgasbord of avenues that can be explored that cultivate a strong synergy between astrodynamical considerations and observations using space VLBI. For example, the impact/advantage of $J_{2}$ effect for making polarimetric observations of the photon ring using Earth-Space VLBI \citep{Palumbo:2023} can be explored that can highlight whether an informed choice of orbital parameters can lead to more robust observables in the presence of astrodynamical perturbations. In a similar vein, one can investigate the similar impact on closure quantities in space VLBI operations, which in our nomenclature would correspond to one chief and at least two deputy satellites for closure amplitudes and three deputies for closure phases. Herein, one can look at how a cluster of satellites (say 3), while operating under the $J_{2}$ effect, can lead to robust interferometric closure observables; the appropriate orbit selection for these has been studied in \citet{Marsden2001} and formation of orbits of such clusters including the $J_{2}$ perturbation effect has also been investigated \citet{Sam2002}. We note that the $J_{2}$ effect is even more important here since the most significant error when modelling relative orbits of a chief and deputy is due to the assumption that the Earth is spherical \citep{Sam2002}.\\
In terms of developments focusing on developing an astrodynamical framework tailored to VLBI, one can envision a ``dynamic" space-VLBI mission in which we utilise a suite of astrodynamical manoeuvres that can adjust the orbit parameters of the orbiter so as to optimise its coverage for several sources over a given mission cycle. For example, when an orbiter forms a baseline with an Earth-based station, the semi-major axis of the orbiter can be increased or decreased so as to investigate photon ring science and jet related structures respectively. Manoeuvres such as changing individual orbital parameters such as inclination, RAAN etc can also be performed to optimise pointing at a particular source of interest.  A comprehensive discussion of such manoeuvres is given in \citet{Vallado:2006} and one can investigate whether taking advantage of the $J_{2}$ effect in the spirit of the discussion provided in the paper can help reduce the fuel budget for such missions. We expect to explore these avenues in the near future.\\
Lastly, the results of this paper suggest that an ambitious mission in the distant future of having an ``ALMA-like" configuration in space can be considered wherein the orbit selection informed by the $J_{2}$ effect can provide extremely dense $(u,v)$ coverage over shorter baselines. Such orbiters, can also work in conjunction with other stations at faraway points like $L2$ to have very long baselines and thereby catering to observations of the black hole photon ring. This deployment can provide a VLBI network in space that is in principle similar to how ALMA currently operates with Earth-based stations to cater to a variety of scientific goals. We hope that the results of this paper would stimulate investigations into expanding the scientific scope of radio astronomy missions from space. 

\section{acknowledgments}
A.T. acknowledges fruitful discussions  with Kazunori Akiyama, Vincent Fish, Soung Sub Lee and Sascha Trippe. D.C.M.P. was supported by the Black Hole Initiative at Harvard University, which is funded by grants from the John Templeton Foundation (JTF-62286) and the Gordon and Betty Moore Foundation (GBMF-8273.01). 

%

\vspace{5mm}




\appendix
\section[The J2 Equations of Motion]{The $J_{2}$ Equations of Motion}
The general equation of motion under the influence of the $J_{2}$ effect is given by:
\begin{gather}
     \ddot{\vec{r}}=-\frac{\mu}{r^{2}}+J_{2}(\vec{r})\equiv g(\vec{r})+J_{2}(\vec{r}),\label{eq:geneom}
\end{gather}
where
\begin{gather}
    J_{2}(\vec{r})=-\frac{3J_{2}\mu R_{e}^{2}}{2r^{4}}[(1-3\sin^{2}i\sin^{2}\theta)\hat{x}+(2\sin^{2}i\sin\theta\cos\theta)\hat{y}+(2\sin i\cos i\sin\theta)\hat{z}].
\end{gather}
where $\hat{x}$ points in the radial direction of the satellite's position vector, $\hat{z}$ is perpendicular to the orbital plane and $\hat{y}$ completes the orthogonal triad and points in the direction of motion of the satellite. For all our cases, we would consider a circular reference orbit of the chief ($r_{\text{ref}}\equiv r_{c}$) under the influence of the gravitational potential of the spherical Earth, and so the equation of motion of the reference orbit is given by:
\begin{gather}
    \ddot{\vec{r}}_{ref}=g(\vec{r}_{ref}),
\end{gather}
where $g(\vec{r})$ is the gravitational force due to a spherical Earth and the subscript $\text{ref}$ implies that it is the force experienced by the chief in its frame. The equation of motion of the deputy is obtained by linearising the gravitational terms in \ref{eq:geneom} with respect to the reference orbit:
\begin{gather}
    \ddot{\vec{r}}=g(\vec{r})+\nabla g(\vec{r}).(\vec{r}-\vec{r}_{\text{ref}})+J_{2}(\vec{r}_{\text{ref}})+\nabla J_{2}(\vec{r}_{\text{ref}}).(\vec{r}-\vec{r}_{\text{ref}}), \label{eq:lingrav}
\end{gather}
where the ``dot" implies dot product between the vectors. The gradient terms in the $(r-\theta-i)$ system are given by:
\begin{gather}
    \nabla g(\vec{r})=\begin{pmatrix}
        \frac{2\mu}{r}&0&0\\0&\frac{-\mu}{r}&0\\0&0&\frac{-\mu}{r}        \end{pmatrix}, \nonumber \\
        \nabla J_{2}=\frac{6\mu J_{2}R_{e}^{2}}{r_{\text{ref}}^{2}}\begin{pmatrix}
            (1-3\sin^{2}i\sin^{2}\theta)&\sin^{2}i\sin2\theta&\sin 2i\sin\theta\\ \sin^{2}i\sin2\theta&-\frac{1}{2}-\sin^{2}i(\frac{1}{2}-\frac{7}{4}\sin^{2}\theta)&\frac{-\sin 2i\cos\theta}{4}\\ \sin 2i\sin\theta&\frac{-\sin2i\cos\theta}{4}&-\frac{3}{4}-\sin^{2}i(\frac{1}{2}+\frac{5}{4}\sin^{2}\theta)
    \end{pmatrix}.
\end{gather}
Now, taking the motion with respect to the reference orbit, we have
\begin{gather}
    \vec{x}=\vec{r}-\vec{r}_{\text{ref}},
\end{gather}
and since the reference orbit is rotating, with say angular velocity $\omega$, the equation of motion, obtained by taking the double derivative of $\vec{x}$ will have contributions from $\omega$. From the known transformation in classical dynamics to a rotating co-ordinate system, we get:
\begin{gather}
    \ddot{\vec{x}}=\ddot{\vec{r}}-\ddot{\vec{r}}_{ref}-2\vec{\omega}\times\dot{\vec{x}}-\dot{\vec{\omega}}\times\vec{x}-\vec{\omega}\times(\vec{\omega}\times\vec{x}). \label{eq:rotcs}
\end{gather}
For a circular reference orbit, 
\begin{gather}
    \vec{\omega}=\sqrt{\frac{\mu}{r_{\text{ref}}^{3}}}\hat{z}\equiv n\hat{z}.
\end{gather}
Finally, using Equation \ref{eq:lingrav} into Equation \ref{eq:rotcs}, we get:
\begin{gather}
\ddot{\vec{x}}+2\vec{\omega}\times\dot{\vec{x}}+\dot{\vec{\omega}}\times\vec{x}+\vec{\omega}\times(\vec{\omega}\times\vec{x})=g(\vec{r}_{\text{ref}})+\nabla g(\vec{r}_{\text{ref}}).\vec{x}+J_{2}(\vec{r}_{\text{ref}})+\nabla J_{2}(\vec{r}_{\text{ref}}).\vec{x}-\ddot{r}_{\text{ref}}. \label{eq:eomj2}
\end{gather}

Next, we take into account further considerations that arise due to the presence of the $J_{2}$ dependent terms in Equation \ref{eq:eomj2}. The equation is a linearised equation of motion in which the term $\nabla J_{2}(\vec{r}_{ref})$ is not constant except for equatorial orbits. An approximate solution to take this into account is to time average the term:
\begin{gather}
    \frac{1}{2\pi}\int_{0}^{2\pi}\nabla J_{r}(\vec{r})d\theta=\frac{\mu}{r^{3}}\begin{pmatrix}
        4s&0&0\\0&-s&0\\0&0&-3s
    \end{pmatrix}, \nonumber \\ s=\frac{3J_{2}R_{e}^{2}}{8r^{2}}(1+3\cos2i).
\end{gather}
Next, the effect of the $J_{2}$ force is that the perturbed satellite has a different period when compared to the case when the $J_{2}$ effect was absent. Due to this, the deputy drifts apart from the reference orbit and the linearised equations break down. To fix this, we take the time average of the ``$J_{2}$-force" term as well:
\begin{gather}
    \frac{1}{2\pi}\int_{0}^{2\pi}J_{2}(\vec{r})d\theta=-n^{2}rs\hat{x}.
\end{gather}
This changes the reference orbit,
\begin{gather}
    \ddot{\vec{r}}_{\text{ref}}=g(\vec{r}_{\text{ref}})+\frac{1}{2\pi}\int_{0}^{2\pi}J_{2}(\vec{r}_{ref})d\theta, \label{eq:reforb1}
\end{gather}
which in turn changes the angular velocity $\omega$ of the rotating co-ordinate system. The new velocity can be found from
\begin{gather}
\vec{\omega}\times(\vec{\omega}\times\vec{r}_{ref})=g(\vec{r}_{ref})+\frac{1}{2\pi}\int_{0}^{2\pi}J_{2}(\vec{r}_{ref})d\theta,
\end{gather}
that gives the new angular velocity as:
\begin{gather}
    \vec{\omega}=n\sqrt{1+s}\hat{z}\equiv nc\hat{z}.
\end{gather}
Thus, our final equation of motion that incorporates all these effects is given by:
\begin{gather}
\ddot{\vec{x}}+2\vec{\omega}\times\dot{\vec{x}}+\dot{\vec{\omega}}\times\vec{x}+\vec{\omega}\times(\vec{\omega}\times\vec{x})=\nabla g(\vec{r}_{ref}).\vec{x}+J_{2}(\vec{r}_{ref})\nonumber \\ +\frac{1}{2\pi}\int_{0}^{2\pi}\nabla J_{2}(\vec{r}_{ref})d\theta.\vec{x}-\frac{1}{2\pi}\int_{0}^{2\pi}J_{2}(\vec{r}_{ref})d\theta. \label{eq:eomschweig2}
\end{gather}

\subsection{Case II.A: including cross-track drift}
As found by Schweighart, the solution to the equation of motion \ref{eq:eomschweig2} for the components $\vec{x}=(x,y,z)$ of the relative separation vector is given by:
\begin{gather}
    x(t)=(x_{0}-\alpha)\cos(nt\sqrt{1-s})+\frac{\sqrt{1-s}}{2\sqrt{1+s}}y_{0}\sin(nt\sqrt{1-s})+\alpha\cos(2nt\sqrt{1+s}),\\
    y(t)=-\frac{2\sqrt{1+s}}{\sqrt{1-s}}(x_{0}-\alpha)\sin(nt\sqrt{1-s})+y_{0}\cos(nt\sqrt{1-s})+\frac{1+3s}{2(1+s)}\alpha\sin(2nt\sqrt{1+s}),\\
    z(t)=z_{0}\cos(nt\sqrt{1+3s})+\frac{\dot{z_{0}}}{n\sqrt{1+3s}}\sin(nt\sqrt{1+3s})\nonumber\\ +\beta\Big(\sqrt{1+s}\sin(nt\sqrt{1+3s})-\sqrt{1+3s}\sin(nt\sqrt{1+s})\Big),
\end{gather}
wherein
\begin{gather}
    \alpha=\frac{3J_{2}R_{e}^{2}}{8r_{\text{ref}}(3+5s)}(1-\cos2i_{\text{ref}}),\quad\beta=\frac{3J_{2}R_{e}^{2}\sin2i_{\text{ref}}}{4r_{\text{ref}}s\sqrt{1+3s}},
\end{gather}
and $x_{0},y_{0},z_{0}$ are the initial positions and $\dot{x}_{0},\dot{y}_{0}$ and $\dot{z}_{0}$ are the initial velocities. For the case of no drift and offset in any direction, we have:
\begin{gather}
    \dot{x}_{0}=y_{0}n\Big(\frac{1-s}{2c}\Big),\quad y_{0}=-2ncx_{0}+\frac{3n J_{2}R_{e}^{2}(1-\cos2i_{\text{ref}})}{8k r_{\text{ref}}}.
\end{gather}
Note that there is an error in Schweighart's Eq. 3.27 \citep{Schweig} in that the term in the denominator of $\dot{y}_{0}$ has to be $8k$ instead of $8c$.\\
In the equations above, even though the $J_{2}$ effect has been incorporated, the orbiter still drifts apart from the reference orbit in the $z$ direction. This can be seen as follows. Suppose we set the initial conditions to be 
\begin{gather}
    x_{0}=0\text{(km)}, \quad y_{0}=0\text{(km)}\quad z_{0}=0\text{(km)}, \nonumber\\
\dot{x}_{0}=0(\text{(km/s)}),\quad \dot{y}_{0}=0(\text{(km/s)}), \quad \dot{z}_{0}=0(\text{(km/s)}).
\end{gather}
Then,the equation of motion in $z$ direction is
\begin{gather}
z(t)=\beta\Big(\sqrt{1+s}\sin(nt\sqrt{1+3s})-\sqrt{1+3s}\sin(nt\sqrt{1+s})\Big).
\end{gather}

\begin{figure}[h]
    \centering
    \includegraphics[width=0.7\textwidth]{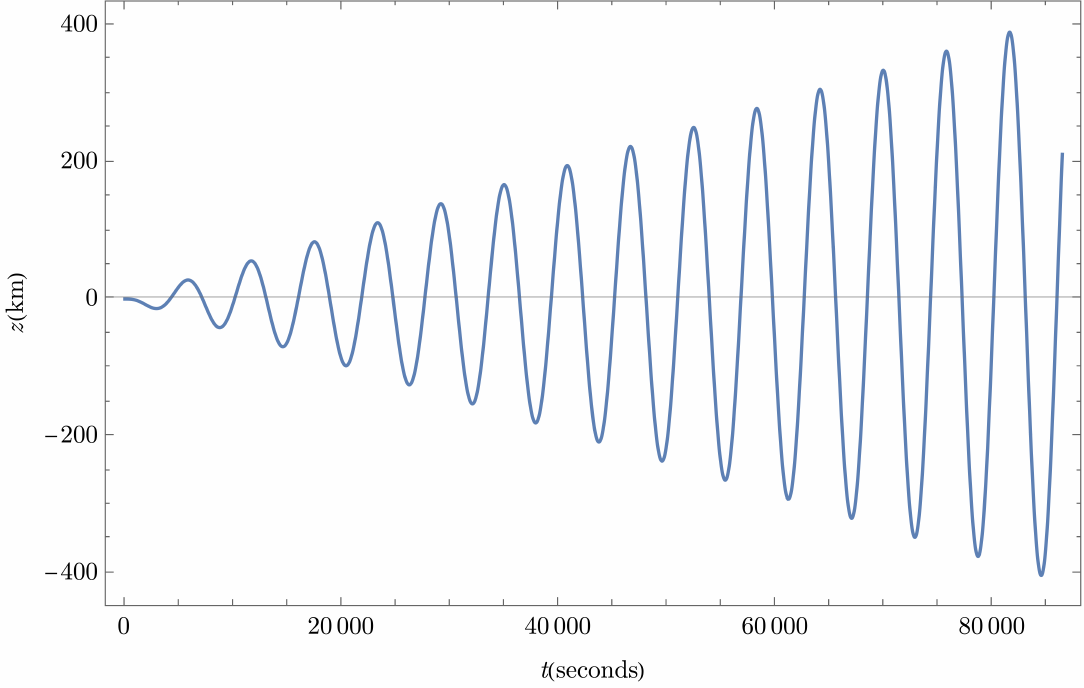}
    \caption{The drift in $z$ direction from Schweighart's equations.}
    \label{f:driftzj2}
\end{figure}

For the orbital parameters of the chief given in \ref{tab:orbchiefzdrift}, the drift in the z-direction is apparent in the Figure \ref{f:driftzj2}.
\begin{table}[h]
\begin{center}
$\begin{array}{cc}
    \begin{tabular}{m{2cm} m{2cm}}
        \toprule
        Orb. Param. & Value   \\
        \midrule
        $a_{c}$(km)	& $7000$  \\
        $e_{c}$	& $0$   \\
        $i_{c}(^\circ)$ & $35$ \\
        $\theta_{c}(^\circ)$	& $0$   \\
        $\Omega_{c}(^\circ)$ &  $0$    \\
        \bottomrule
    \end{tabular}
    \end{array}$
    \end{center}
    \caption{Orbital Parameters of the Chief for demonstrating drift in z-direction.}
    \label{tab:orbchiefzdrift}
\end{table}

For these given values, since $z(t)\neq0$, the expression for $v(\lambda)$ contains a contribution from the $z(t)$ component. Thus, it fails to have stable, bounded motion thereby making it the $(u,v)$ plane devoid of coherent tracks. A solution suitable for physical applications that attempts to fix this is given in the next subsection.

\subsection{Case II.B: correcting cross-track drift}
\citet{Schweig} identified that the normal component of the $J_{2}$ force causes the drift in the $z$-direction. This affects four out of the six orbital elements and by solving the equations that govern their time evolution, under the assumption of constant inclination, the time evolution of the orbital elements is given by \citep{Schweig}:
\begin{gather}
    i(t)=i_{0}-\frac{3\sqrt{\mu}J_{2}R_{e}^{2}}{2kr^{7/2}}\cos i\sin i\sin^{2}(2kt),\\
    \Omega(t)=\Omega_{0}-\frac{3\sqrt{\mu}J_{2}R_{e}^{2}}{2kr^{7/2}}\cos i\Bigg[tk-\frac{\sin(2kt)}{2}\Bigg],\\
    \theta(t)=nct+\frac{3\sqrt{\mu}J_{2}R_{e}^{2}}{2kr^{7/2}}\cos^{2}i\Bigg[tk-\frac{\sin(2kt)}{2}\Bigg],
\end{gather}
where
\begin{gather}
k=nc+\frac{3\sqrt{\mu}J_{2}R_{e}^{2}}{2kr^{7/2}}.
\end{gather}
To fix the resultant drift, Schweighart added this component to the reference orbit in \ref{eq:reforb1} as follows:
\begin{gather}
    \ddot{\vec{r}}_{ref}=g(\vec{r}_{ref})+\frac{1}{2\pi}\int_{0}^{2\pi}J_{2}(\vec{r}_{ref})d\theta+\vec{J}_{2}(\vec{r}_{ref}).\hat{N}.
\end{gather}
Now, both the perturbed and the reference orbiter have a drift in the longitude of the ascending node and thus the satellites will not drift apart.\\
The solution to the components $\vec{x}=(x,y,z)$ in this case is given by:
\begin{gather}
    x(t)=(x_{0}-\alpha_{2})\cos(nt\sqrt{1-s})+\frac{\sqrt{1-s}}{2\sqrt{1+s}}\dot{y}_{0}\sin(nt\sqrt{1-s})+\alpha_{2}\cos(2kt),\\
    y(t)=-\frac{2\sqrt{1+s}}{\sqrt{1-s}}(x_{0}-\alpha_{2})\sin(nt\sqrt{1-s})+\dot{y}_{0}\cos(nt\sqrt{1-s})+\beta_{2}\sin(2kt),\\
    z(t)=z_{0}\cos(nt\sqrt{1+3s})+\frac{\dot{z}_{0}}{n\sqrt{1+3s}}\sin(nt\sqrt{1+3s}),
\end{gather}
wherein
\begin{gather}
    \alpha_{2}=-\frac{3J_{2}R_{e}^{2}n^{2}(3k-2n\sqrt{1+s}}{8kr_{ref}(n^{2}(1-s)-4k^{2})}(1-\cos 2i),\\
    \beta_{2}=-\frac{3J_{2}R_{e}^{2}n^{2}(2k(2k-3n\sqrt{1+s})+n^{2}(3+5s))}{2k(n^{2}(1-s)-4k^{2})}{2k(n^{2}(1-s)-4k^{2})}(1-\cos 2i).
\end{gather}
All other parameters have the same values as the previous case. Note that there is another (rather non-trivial) error in Schweighart's equations. Namely the equations for $x$ and $y$ were incorrectly written with $y_{0}$ instead of the corrected expression above that has $\dot{y}_{0}$. This can be double checked with \citet{Ginn}. Curiously, the resultant plots made by Schweighart use the correct expressions (i.e using $\dot{y}_{0}$ instead of $y_{0}$).

\section{Notation and Terminology}
Since the results of the paper rely on the cross-fertilisation of astrodynamical and VLBI considerations, an accessible tabulation of the related terminology has been provided in Table \ref{tab:terminology}.
\begin{table}[h!]
\begin{center}
$\begin{array}{cc}
    \begin{tabular}{m{6cm}| m{6cm}}
        \toprule
        Terminology & Description \\
        \hline
        \midrule
        Argument of Perigee ($\omega$) & Orientation of the ellipse in the orbital plane (measured from the ascending node to the periapsis).\\ \hline
     
        Baseline & Vector drawn between two telescopes observing the same source orthographically projected to the source.\\ \hline
    
        Chief Orbiter & The primary orbiter in reference to which the motion of other orbiters is studied.\\ \hline
  
        Deputy Orbiter & The orbiter moving relative to the primary orbiter. There can be several deputy orbiters moving in reference to a primary orbiter.\\ \hline
     
        Eccentricity (\(e\))  & Elliptical shape of the orbit (\(0<e<1\)). \\ \hline
     
        Earth Centered Intertial (ECI) Frame & Earth-centered coordinate system fixed with respect to the celestial sphere.\\ \hline
       
        Local Vertical Local Horizontal (LVLH) frame & Spacecraft-centered coordinate system with respect to the nadir direction and the perpendicular local horizontal.\\ \hline
        
        Hour Angle and Declination ($H,\delta$) & The standard co-ordinates used to locate the position of an astronomical source on the celestial sphere.\\ \hline
       
        Inclination (\(i\)) & Orientation of the orbit with respect to the equator.\\ \hline
        
        $J_{2}$ effect & The effect on the dynamical motion of the body arising due to the oblateness of the Earth.\\ \hline
        
        Mean Anomaly $M$ & Fraction of an elliptical orbit's period that has elapsed since passing the periapsis. \\ \hline
       
        Right Ascension of Ascending Node $(\Omega)$ & Intersection of the ascending direction of the orbit and the equator, with respect to the vernal equinox.\\ \hline
   
        Semi-Major Axis $(a)$ & Size of the orbit (average of the apoapsis and periapsis radii).\\ \hline
   
        $(u,v)$ coverage & The set of image-conjugate Fourier coefficients sampled by the baselines swept out by a set of telescopes during a given observations. \\ \hline

        (3-1-3) Euler angle system & Orientation of a rigid body with respect to an inertial coordinate system, described by three rotation angles such that the order of rotation is done first for the ``third" axis of the original system, then the ``first" axis of the intermediate  system and then finally on the``third" axis of the final transformed system.  \\ \hline
    \end{tabular}

\end{array}$
\end{center}
\caption{Terminology used in the paper}
    \label{tab:terminology}
\end{table}
\clearpage
\bibliography{sample631}{}



\end{document}